\documentclass[aip,jcp,reprint,noshowkeys,10pt]{revtex4-1}
\usepackage{graphicx,dcolumn,bm,xcolor,microtype,multirow,amscd,amsmath,amssymb,amsfonts,physics,mhchem,longtable,wrapfig,xspace}

\usepackage{natbib}
\AtBeginDocument{\nocite{achemso-control}}

\usepackage{mathpazo,libertine}

\usepackage{listings}
\usepackage[lined,boxed,commentsnumbered]{algorithm2e}

\SetKw{KwBy}{by} 

\SetCommentSty{mycommfont}

\usepackage[normalem]{ulem}

\definecolor{darkgreen}{RGB}{0, 180, 0}

\usepackage{hyperref}
\hypersetup{
    colorlinks=true,
    linkcolor=blue,
    filecolor=blue,      
    urlcolor=blue,	
    citecolor=blue
}
\newcommand{\mU}{\mathbf{U}}
\newcommand{\mS}{\mathbf{S}}
\newcommand{\mW}{\mathbf{W}}
\newcommand{\mR}{\mathbf{R}}
\newcommand{\mh}{\mathbf{h}}
\newcommand{\my}{\mathbf{y}}
\newcommand{\mC}{\mathbf{C}}
\newcommand{\mI}{\mathbf{I}}

\newcommand{\mcI}{\mathcal{I}}
\newcommand{\mcA}{\mathcal{A}}
\newcommand{\mcAs}{\mathcal{A}^{\star}}
\newcommand{\mcD}{\mathcal{D}}
\newcommand{\mcS}{\mathcal{S}}

\newcommand{\la}{\lambda}
\newcommand{\si}{\sigma}
\newcommand{\eps}{\varepsilon}

\newcommand{\PsiO}{\Psi^{(0)}}
\newcommand{\kPsiO}{\ket*{\Psi^{(0)}}}
\newcommand{\ePT}[1]{e_{#1}^{(2)}}
\newcommand{\tePT}[1]{\tilde{e}_#1^{(2)}}

\newcommand{\Hij}[3][\hH]{\mel*{#2}{#1}{#3}}
\newcommand{\Sij}[3][\hS^2]{\mel*{#2}{#1}{#3}}

\newcommand{\Norb}{N_\text{orb}}
\newcommand{\Nst}{N_\text{states}}
\newcommand{\Nnodes}{N_\text{nodes}}
\newcommand{\Ndet}{N_\text{det}}

\newcommand{\Nelec}{N}
\newcommand{\Nalpha}{N_{\uparrow}}
\newcommand{\Nbeta}{N_{\downarrow}}
\newcommand{\Nint}{N_\text{int}}

\newcommand{\Nalphadet}{{N_\text{det}^{\uparrow}}}
\newcommand{\Nbetadet}{{N_\text{det}^{\downarrow}}}

\newcommand{\bit}[1]{{{#1}}}
\newcommand{\bitP}{{\bit{P}}}
\newcommand{\bitI}{{\bit{I}}}

\newcommand{\bitx}[2]{{{#1}_{#2}}}
\newcommand{\bitIsigma}{{\bitx{I}{\sigma}}}
\newcommand{\bitPsigma}{{\bitx{P}{\sigma}}}

\newcommand{\FALSE}{{\text{\texttt{FALSE}}}}

\newcommand\QP{\textsc{Quantum Package}\xspace}

\newcommand{\EHF}{E_\text{HF}}
\newcommand{\Ec}{E_\text{c}}
\newcommand{\Eex}{E_\text{exact}}

\newcommand{\EFCI}{E_\text{FCI}}
\newcommand{\EPT}{E^{(2)}}
\newcommand{\SigPT}{\Sigma^{(2)}}
\newcommand{\EO}{E^{(0)}}

\newcommand{\cMO}[1]{C_{#1}}
\newcommand{\cCI}[1]{c_{#1}}

\newcommand{\AO}[1]{\chi_{#1}}
\newcommand{\MO}[1]{\phi_{#1}}

\newcommand{\vac}{ {\ket{}}}
\newcommand{\ac}[1]{a^\dagger_{#1}}

\newcommand{\hH}{\Hat{H}}

\newcommand{\hh}{\Hat{h}}
\newcommand{\hI}{\Hat{I}}
\newcommand{\hS}{\Hat{S}}
\newcommand{\ordering}{{\hat{\mathcal{O}}}}

\newcommand{\kalpha}{{\ket{\alpha}}}
\newcommand{\kalphas}{{\ket{\alpha^{\star}}}}

\newcommand{\kI}{\ket{I}}
\newcommand{\bI}{\bra{I}}
\newcommand{\kJ}{\ket{J}}

\newcommand{\Ipqrs}{I^{rs}_{pq}}
\newcommand{\kIpqrs}{\ket*{I^{rs}_{pq}}}

\newcommand{\br}{{\mathbf{r}}}

\newcommand{\popcnt}[1]{\norm{#1}}

\newcommand{\shiftl}[2]{\qty(#1 \ll #2)}

\newcommand{\iand}[2]{#1 \wedge #2}
\newcommand{\ieor}[2]{#1 \oplus #2}

\newcommand{\mH}{\mathbf{H}}

\newcommand{\tabc}[1]{\multicolumn{1}{c}{#1}}

\newcommand{\InAU}[1]{#1 a.u.}

\newcommand{\LCPQ}{Laboratoire de Chimie et Physique Quantiques (UMR 5626), Universit\'e de Toulouse, CNRS, UPS, France}
\newcommand{\LCT}{Laboratoire de Chimie Th\'eorique, Sorbonne Universit\'e, CNRS, Paris, France}
\newcommand{\CSD}{Computational Science Division, Argonne National Laboratory, Argonne, IL 60439, United States}
\newcommand{\CALMIP}{CALMIP, Universit\'e de Toulouse, CNRS, INPT, INSA, UPS, UMS 3667, France}
\newcommand{\ISCD}{Institut des Sciences du Calcul et des Donn\'ees, Sorbonne Universit\'e, F-75005 Paris, France}
\newcommand{\UPitt}{Department of Chemistry, University of Pittsburgh, Pittsburgh, PA 15260, United States}

\begin{document}	

\title{Quantum Package 2.0: An Open-Source Determinant-Driven Suite of Programs}

\author{Yann Garniron}
\affiliation{\LCPQ}
\author{Thomas Applencourt}
\affiliation{\CSD}
\author{Kevin Gasperich}
\affiliation{\CSD}
\affiliation{\UPitt}
\author{Anouar Benali}
\affiliation{\CSD}
\author{Anthony Fert\'e}
\affiliation{\LCT}
\author{Julien Paquier}
\affiliation{\LCT}
\author{Barth\'el\'emy Pradines}
\affiliation{\LCT}
\affiliation{\ISCD}
\author{Roland Assaraf}
\affiliation{\LCT}
\author{Peter Reinhardt}
\affiliation{\LCT}
\author{Julien Toulouse}
\affiliation{\LCT}
\author{Pierrette Barbaresco}
\affiliation{\CALMIP}
\author{Nicolas Renon}
\affiliation{\CALMIP}
\author{Gr\'egoire David}
\affiliation{Aix-Marseille Univ, CNRS, ICR, Marseille, France}
\author{Jean-Paul Malrieu}
\affiliation{\LCPQ}
\author{Micka\"el V\'eril}
\affiliation{\LCPQ}
\author{Michel Caffarel}
\affiliation{\LCPQ}
\author{Pierre-Fran\c{c}ois Loos}
\email{loos@irsamc.ups-tlse.fr}
\affiliation{\LCPQ}
\author{Emmanuel Giner}
\email{emmanuel.giner@lct.jussieu.fr}
\affiliation{\LCT}
\author{Anthony Scemama}
\email{scemama@irsamc.ups-tlse.fr}
\affiliation{\LCPQ}

\begin{abstract}
\begin{wrapfigure}[13]{O}[-1.5cm]{0.35\textwidth}
	\centering
	\includegraphics[width=\linewidth]{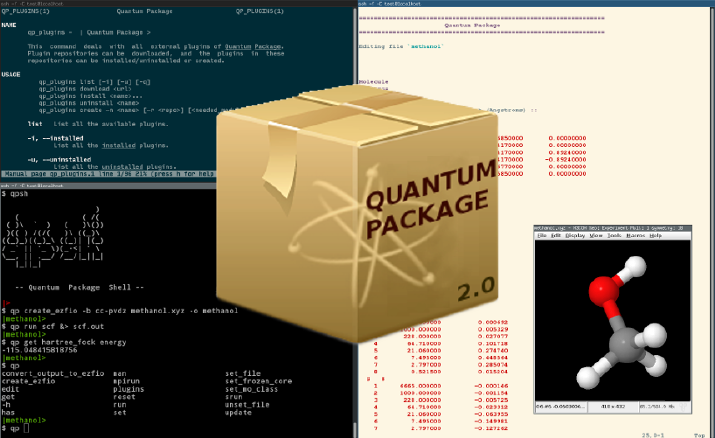}
	\\
	\bf TOC graphical abstract
\end{wrapfigure}
Quantum chemistry is a discipline which relies heavily on very expensive numerical computations.
The scaling of correlated wave function methods lies, in their standard implementation, between $\order*{N^5}$ and $\order*{e^{N}}$, where $N$ is proportional to the system size.
Therefore, performing accurate calculations on chemically meaningful systems requires i) approximations that can lower the computational scaling, and ii) efficient implementations that take advantage of modern massively parallel architectures.
\QP is an open-source programming environment for quantum chemistry specially designed for wave function methods.
Its main goal is the development of determinant-driven selected configuration interaction (sCI) methods and multi-reference second-order perturbation theory (PT2).
The determinant-driven framework allows the programmer to include any arbitrary set of determinants in the reference space, hence providing greater methodological freedom. 
The sCI method implemented in \QP is based on the CIPSI (Configuration Interaction using a Perturbative Selection made Iteratively) algorithm which complements the variational sCI energy with a PT2 correction.
Additional external plugins have been recently added to perform calculations with multireference coupled cluster theory and range-separated density-functional theory. 
All the programs are developed with the IRPF90 code generator, which simplifies collaborative work and the development of new features.
\QP strives to allow easy implementation and experimentation of new methods, while making parallel computation as simple and efficient as possible on modern supercomputer architectures.
Currently, the code enables, routinely, to realize runs on roughly 2\,000 CPU cores, with tens of millions of determinants in the reference space.
Moreover, we have been able to push up to 12\,288 cores in order to test its parallel efficiency.
In the present manuscript, we also introduce some key new developments: i) a renormalized second-order perturbative correction for efficient extrapolation to the full CI limit, and ii) a stochastic version of the CIPSI selection performed simultaneously to the PT2 calculation at no extra cost.
\end{abstract}

\maketitle


\section{Introduction}
\label{sec:intro}

In 1965, Gordon Moore predicted that the number of transistors in an integrated circuit would double about every two years (the so-called Moore's law). \cite{Moore_1965}  
Rapidly, this ``law'' was interpreted as an expected two-fold increase in performance every 18 months.
This became an industrial goal.
The development of today's most popular electronic structure codes was initiated in the 1990's (or even before).  
At that time, the increase of computational power from one supercomputer generation to the next was mostly driven by an increase of processors' frequency.
Indeed, the amount of random access memory was small, the time to access data from disk was slow, and the energy consumption of the most powerful computer was 236~kW, hence far from being an economical concern. \cite{top500_93}
At the very beginning of the 21st century, having increased continuously, both the number of processors and their frequency raised the supercomputer power consumption by two orders of magnitude, inflating accordingly the electricity bill.  
The only way to slow down this frenetic growth of power consumption while keeping alive Moore's dream was to freeze the processor's frequency (between 1 and 4~GHz), and increase the number of CPU cores.  
The consequence of such a choice was that \emph{``free lunch''} was over: the programmers now had to parallelize their programs to make them run faster. \cite{Sutter_2005}
At the same time, computer scientists realized that the increase of performance in memory access was slower than the increase in computational power, \cite{Wulf_1995} and that the floating-point operation (or flop) count would soon stop being the bottleneck.
From now on, data movement would be the main concern. 
This paradigm shift was named the \emph{memory wall}. 
Moore's law is definitely near the end of its life. \cite{Khan2018Jan}

The traditional sequential algorithms of quantum chemistry are currently being redesigned and replaced by parallel equivalents by multiple groups around the world. \cite{Booth_2009, Booth_2010, Cleland_2010, Sharma_2017, Garniron_2017b, Smith_2017, Neuhauser_2013, Willow_2014, Willow_2012, Johnson_2017, Johnson_2016, Gruneis_2017, Doran_2016}
This has obviously a significant influence on methodological developments.
The most iconic example of this move towards parallel-friendly methods is the recently developed \emph{full configuration interaction quantum Monte Carlo} (FCIQMC) method by Alavi and coworkers. \cite{Booth_2009} 
FCIQMC can be interpreted as a Monte Carlo equivalent of older selected configuration interaction (sCI) algorithms \cite{Bender_1969,Whitten_1969,Huron_1973,ShBuPeyChemPhys78,BuePeyButMolPhys78,Evangelisti_1983, Cimiraglia_1985, Cimiraglia_1987, Illas_1988, Povill_1992,EngHanLenCompChem01,Abrams_2005,Bunge_2006,MusEngelsJCC06,Bytautas_2009,Giner_2013,Caffarel_2014,Giner_2015,Garniron_2017b,Caffarel_2016a,Caffarel_2016b,Holmes_2016,Sharma_2017,Holmes_2017,Chien_2018,Scemama_2018a,Scemama_2018b,Loos_2018b,Garniron_2018,Evangelista_2014,Schriber_2016,Schriber_2017,Liu_2016,Per_2017,Ohtsuka_2017,Zimmerman_2017,Li_2018,Loos_2019} such as CIPSI (Configuration Interaction using a Perturbative Selection made Iteratively),\cite{Huron_1973} that are iterative and thus \emph{a priori} not well adapted to massively parallel architecture. 
As we shall see here, things turn out differently, and the focus of the present article is to show that sCI methods can be made efficient on modern massively parallel supercomputers.

\QP\cite{QP} is an open-source suite of wave function quantum chemistry methods mainly developed at the \emph{Laboratoire de Chimie et Physique Quantiques} (LCPQ) in Toulouse (France), and the \emph{Laboratoire de Chimie Th\'eorique} (LCT) in Paris.
Its source code is freely available on GitHub at the following address: \url{https://github.com/QuantumPackage/qp2}.
\QP strives to allow easy implementation and experimentation of new methods, while making parallel computation as simple and efficient as possible. 
Accordingly, the initial choice of \QP was to go towards \emph{determinant-driven} algorithms.
Assuming a wave function expressed as a linear combination of determinants, a determinant-driven algorithm essentially implies that the outermost loop runs over determinants.
On the other hand, more traditional \emph{integral-driven} algorithms have their outermost loop running on the two-electron integrals appearing in the expression of the matrix elements in the determinant basis (see Sec.~\ref{sec:MatEl}).
Determinant-driven algorithms allow more flexibility than their integral-driven counterparts, \cite{Povill_1995} but they have been known for years to be less efficient than their integral-driven variant for solving electronic structure problems. 
In high-precision calculations, the number of determinants is larger than the number of integrals, justifying the integral-driven choice.  
However, today's programming standards impose parallelism, and if determinant-driven calculations prove to be better adapted to parallelism, such methods could regain popularity. 
More conventional approaches have also been very successfully parallelized: CCSD(T), \cite{Olson_2007, Kjaergaard_2017} DMRG, \cite{Kantian_2019} GW, \cite{Blase_2018} QMC, \cite{Scemama_2013b, Scemama_2016, Kim_2018} and many others.

\QP was used in numerous applications, in particular to obtain reference ground-state energies \cite{Scemama_2014, Caffarel_2014, Giner_2013, Giner_2015, Caffarel_2016a, Caffarel_2016b} as well as excitation energies \cite{giner_cu_2018,Loos_2018b, Loos_2019} for atomic and molecular systems.
For example, in Ref.~\onlinecite{Loos_2018b}, \QP has been used to compute more than hundred very accurate transition energies for states of various characters (valence, Rydberg, $n \rightarrow \pi^*$, $\pi \rightarrow \pi^*$,  singlet, triplet, \ldots) in 18 small molecules.
The high quality and compactness of the CIPSI wave function was also used for quantum Monte Carlo calculations to characterize the ground state of the water and the \ce{FeS} molecules, \cite{Caffarel_2016b, Scemama_2018a} and obtained highly accurate excitation energies. \cite{Scemama_2018b, Dash_2018, Flores_2018}
Of course, the technical considerations were not the main concern of the different articles that were produced. 
Because the present work focused on the actual implementation of the methods at least as much as on the theory behind them, this article is a perfect opportunity to discuss in depth their implementation.

This manuscript is organized as follows. 
In Sec.~\ref{sec:meth}, we briefly describe the main computational methods implemented in \QP as well as newly developed methods and extrapolation techniques.
Section \ref{sec:implementation} deals with their implementation.
In particular, Sec.~\ref{sec:detdrive} discusses the computation of the Hamiltonian matrix elements using determinant-driven algorithms, while Sec.~\ref{sec:dav} focuses on the acceleration of the Davidson diagonalization, a pivotal point of sCI methods.
In Sec.~\ref{sec:sel}, we focus on the determinant selection step used to build compact wave functions. 
In a nutshell, the principle is to incrementally build a reference wave function by scavenging its external space for determinants that interact with it. 
To make this step more affordable, we designed a new stochastic scheme which selects \emph{on the fly} the more important determinants 
while the second-order perturbative (PT2) energy is computed using a hybrid stochastic-deterministic scheme. \cite{Garniron_2017b}
Therefore, the selection part of this new stochastic CIPSI selection is virtually free as long as one is interested in the second-order perturbative correction, which is crucial in many cases in order to obtain near \emph{full configuration interaction} (FCI) results. 
Section~\ref{sec:cipsi_s2} briefly explains how we produce spin-adapted wave functions, and Sec.~\ref{sec:parallelism} describes parallelism within \QP.
The efficiency of the present algorithms is demonstrated in Sec.~\ref{sec:perf} where illustrative calculations and parallel speedups are reported.
Finally, Sec.~\ref{sec:dev} discusses the development philosophy of \QP as well as other relevant technical details.
Unless otherwise stated, atomic units are used throughout.

\section{Methods}
\label{sec:meth}
\subsection{Generalities}

The correlation energy is defined as \cite{Lowdin_1959}
\begin{equation}
	\Ec = \Eex - \EHF,
\end{equation}
where $\Eex$ and $\EHF$ are, respectively, the exact (non-relativistic) energy and the Hartree-Fock (HF) energy in a complete (one-electron) basis set.

To include electron correlation effects, the wave function associated with the $k$th electronic state, $\ket*{\Psi_k}$, may be expanded in the set of all possible $\Nelec$-electron Slater determinants, $\kI$, built by placing $\Nalpha$ spin-up electrons in $\Norb$ orbitals and $\Nbeta$ spin-down electrons in $\Norb$ orbitals (where $\Nelec = \Nalpha + \Nbeta$).
These so-called molecular orbitals (MOs) are defined as linear combinations of atomic orbitals (AOs)
\begin{equation}
\label{eq:def_mo}
	\MO{p}(\br) = \sum_{\mu}^{\Norb} \cMO{\mu p} \AO{\mu}(\br).
\end{equation}
Note that the MOs are assumed to be real valued in the context of this work. 
The eigenvectors of the Hamiltonian $\hH$ are consequently expressed as linear combinations of Slater determinants, i.e.,
\begin{equation}
	\ket*{\Psi_k} = \sum_{I}^{\Ndet} \cCI{Ik} \kI,
\end{equation}
where $\Ndet$ is the number of determinants.
For sake of conciseness, we will restrict the discussion to the ground state (i.e.~$k=0$) and drop the subscript $k$ accordingly.
Solving the eigenvalue problem in this basis is referred to as FCI and yields, for a given basis set, the exact solution of the Schr\"odinger equation.
Unfortunately, FCI is usually computationally intractable because of its exponential scaling with the size of the system.

\subsection{Matrix elements of the Hamiltonian}
\label{sec:MatEl}

In the $\Nelec$-electron basis of Slater determinants, one expects the matrix elements of $\hH$ to be integrals over $3\Nelec$ dimensions.
However, given the two-electron nature of the Hamiltonian, and because the MOs are orthonormal, Slater determinants that differ by more than two spinorbitals yield a zero matrix element.
The remaining elements can be expressed as sums of integrals over one- or two-electron coordinates, which can be computed at a reasonable cost. 
These simplifications are known as Slater-Condon's rules, and reads
\begin{subequations}
\begin{align}
	\Hij{I}{I} 				& = \sum_{i\in \kI} (i|\hh|i) + \frac{1}{2} \sum_{(i,j)\in \kI} (ii||jj),
	\\
	\Hij{I}{I_p^r} 			& = (p|\hh|r) + \sum_{i\in \kI} (pr||ii),
	\\
	\Hij{I}{I_{pq}^{rs}} 	& = (pr||qs),
\end{align}
\end{subequations}
where $\hh$ is the one-electron part of the Hamiltonian (including kinetic energy and electron-nucleus attraction operators),
\begin{equation}
	(p|\hh|q) = \int \MO{p}(\br) \hh(\br) \MO{q}(\br) d\br
\end{equation}
are one-electron integrals, $i \in \kI$ means that $\MO{i}$ belongs to the
Slater determinant $\kI$,  $\ket*{I_{p}^{r}}$ and $\ket*{I_{pq}^{rs}}$ are determinants obtained from $\kI$ by substituting orbitals $\MO{p}$ by $\MO{r}$,  and  $\MO{p}$ and $\MO{q}$ by $\MO{r}$ and $\MO{s}$, respectively, 
\begin{equation}
	(pq|rs) = \iint \MO{p}(\br_1)\MO{q}(\br_1) r_{12}^{-1} \MO{r}(\br_2)\MO{s}(\br_2)d\br_1 d\br_2
\end{equation}
are two-electron electron repulsion integrals (ERIs), $r_{12}^{-1} = \abs{\br_1 - \br_2}^{-1}$ is the Coulomb operator, and $(pq||rs) = (pq|rs) - (ps|rq)$ are the usual antisymmetrized two-electron integrals.

Within the HF method, Roothaan's equations allow to solve the problem in the AO basis.\cite{Roothaan_1951}
In this context, one needs to compute the $\order*{\Norb^4}$ two-electron integrals $(\mu\nu|\la\si)$ over the AO basis.
Thanks to a large effort in algorithmic development and implementation,\cite{Obara_1986,Head_Gordon_1988,Tenno_1993,Gill_1989,Gill_1991,Libint2,Barca_2017,Zhang_2018} these integrals can now be computed very fast on modern computers.
However, with post-HF methods, the computation of the two-electron integrals is a potential bottleneck.
Indeed, when computing matrix elements of the Hamiltonian in the basis of Slater determinants, ERIs over MOs are required.
Using Eq.~\eqref{eq:def_mo}, the cost of computing a single integral $(pq|rs)$ scales as $\order*{\Norb^4}$.
A naive computation of all integrals in the MO basis would cost ${\cal O}(\Norb^8)$. Fortunately, computing all of them can be scaled down to ${\cal O}(\Norb^5)$ by transforming the indices one by one.\cite{Wilson_1987}
This step is known as the four-index integral transformation.  
In addition to being very costly, this step is hard to parallelize in a distributed way, because it requires multiple collective communications.\cite{Rajbhandari_2017,Limaye_1994,Fletcher_1999,Covick_1990}
However, techniques such as density fitting (also called the resolution of the identity), \cite{Whitten_1973,Eichkorn_1995,Schmitz_2017} low-rank approximations, \cite{Beebe_1977,Aquilante_2007,Roeggen_2008,Peng_2017} or the combination of both \cite{Pham_2019} are now routinely employed to overcome the computational and storage bottlenecks.

\subsection{Selected CI methods}
The sCI methods rely on the same principle as the usual configuration interaction (CI) approaches, except that determinants are not chosen \emph{a priori} based on occupation or excitation criteria, but selected among the entire set of determinants based on their estimated contribution to the FCI wave function. 
Indeed, it has been noticed long ago that, even inside a predefined subspace of determinants, only a small number of them significantly contributes.\cite{Bytautas_2009,Anderson_2018} 
Therefore, an \emph{on-the-fly} selection of determinants is a rather natural idea that has been proposed in the late 1960's by Bender and Davidson \cite{Bender_1969} as well as Whitten and Hackmeyer.\cite{Whitten_1969}
sCI methods are still very much under active development.
The main advantage of sCI methods is that no \emph{a priori} assumption is made on the type of electronic correlation. 
Therefore, at the price of a brute force calculation, a sCI calculation is less biased by the user's appreciation of the problem's complexity.

The approach that we have implemented in \QP is based on the CIPSI algorithm developed by Huron, Rancurel and Malrieu in 1973, \cite{Huron_1973} that iteratively selects external determinants $\kalpha$ --- determinants which are not present in the (reference or variational) zeroth-order wave function 
\begin{equation}
	\kPsiO = \sum_I \cCI{I} \kI
\label{eq:Psivar}
\end{equation}
at a given iteration --- using a perturbative criterion
\begin{equation}
	\label{eq:e2}
	\ePT{\alpha} 
	= \frac{\Hij{\PsiO}{\alpha}^2}{\EO - \Hij{\alpha}{\alpha}},
\end{equation}
where
\begin{equation}
	\label{eq:evar}
        \EO = \frac{\mel*{\PsiO}{\hH}{\PsiO}}{\braket*{\PsiO}{\PsiO}} \ge \EFCI
\end{equation}
is the zeroth-order (variational) energy, and $\ePT{\alpha}$ the (second-order) estimated gain in correlation energy that would be brought by the inclusion of $\kalpha$. 
The second-order perturbative correction 
\begin{equation}
	\label{eq:PT2}
	\EPT 
	= \sum_{\alpha} \ePT{\alpha} 
	= \sum_{\alpha} \frac{\Hij{\alpha}{\PsiO}^2}{\EO - \Hij{\alpha}{\alpha}}
\end{equation}
is an estimate of the total missing correlation energy, i.e., $\EPT \approx \EFCI - \EO$, for large enough expansions.

Let us emphasize that sCI methods can be applied to any determinant space.
Although presented here for the FCI space, it can be trivially generalized to a complete active space (CAS), but also to standard CI spaces such as CIS, CISD or MR-CISD.
The only required modification is to set to zero the contributions associated with the determinants which do not belong to the target space.

There is, however, a computational downside to sCI methods. 
In conventional CI methods, the rule by which determinants are selected is known \emph{a priori}, and therefore, one can map a particular determinant to some row or column indices.\cite{Knowles_1984} 
As a consequence, it can be systematically determined to which matrix element of $\hH$ a two-electron integral contributes.
This allows for the implementation of so-called \emph{integral-driven} methods that work essentially by iterating over integrals.
On the contrary, in (most) sCI methods, the determinants are selected \emph{a posteriori}, and an explicit list has to be maintained as there is no immediate way to know whether or not a determinant has been selected. 
Consequently, we must rely on the so-called \emph{determinant-driven} approach in which iterations are performed over determinants rather than integrals. 
This can be a lot more expensive, since the number of determinants $\Ndet$ is typically much larger than the number of integrals. 
The number of determinants scales as $\order*{\Norb!}$ while the number of integrals scales (formally) as $\order*{\Norb^4}$.
What makes sCI calculations possible in practice is that sCI methods generate relatively compact wave functions, i.e. wave functions where $\Ndet$ is much smaller (by orders of magnitude) than the size of the FCI space.
Furthermore, determinant-driven methods require an effective way to compare determinants in order to extract the corresponding excitation operators, and a way to rapidly fetch the associated integrals involved, as described in Sec.~\ref{sec:detdrive}.

Because of this high computational cost, approximations have been proposed.\cite{Evangelisti_1983} 
Recently, the semi-stochastic heat-bath configuration interaction (SHCI) algorithm has taken further the idea of a more approximate but extremely cheap selection. \cite{Holmes_2016, Sharma_2017,Li_2018} 
Compared to CIPSI, the selection criterion is simplified to
\begin{equation}
	e^{\text{SHCI}}_\alpha = \max_{I} \qty(\abs{c_I \Hij{I}{\alpha}}).
\end{equation}
This algorithmically allows for an extremely fast selection of doubly-excited determinants by an integral-driven approach. Nonetheless, the bottlenecks of the SHCI are  the diagonalization step and the computation of $E^{(2)}$, which remain determinant driven. 

As mentioned above, FCIQMC is an alternative approach of stochastic nature recently developed in Alavi's group, \cite{Booth_2009,Booth_2010,Cleland_2010} where signed walkers spawn from one determinant to connected ones, with a probability that is a function of the associated matrix element. 
The average proportion of walkers on a determinant converges to its coefficient in the FCI wave function.
A more ``brute force'' approach is the purely stochastic selection of Monte Carlo CI (MCCI), \cite{Greer_1995,Greer_1998} where determinants are randomly added to the zeroth-order wave function. 
After diagonalization, the determinants of smaller coefficient are removed, and new random determinants are added.

A number of other variants have been developed but are not detailed here.
\cite{Bender_1969,Whitten_1969,Huron_1973,Evangelisti_1983, Cimiraglia_1985, Cimiraglia_1987, Illas_1988, Povill_1992,Abrams_2005,Bunge_2006,Bytautas_2009,Giner_2013,Caffarel_2014,Giner_2015,Garniron_2017b,Caffarel_2016a,Caffarel_2016b,Holmes_2016,Sharma_2017,Holmes_2017,Chien_2018,Scemama_2018a,Scemama_2018b,Loos_2018b,Garniron_2018,Evangelista_2014,Schriber_2016,Schriber_2017,Liu_2016,Per_2017,Ohtsuka_2017,Zimmerman_2017,Li_2018,Ohtsuka_2017, Coe_2018,Loos_2019}  Although these various approaches appear under diverse acronyms, most of them rely on the very same idea of selecting determinants iteratively according to their contribution to the wave function or energy.

\subsection{Extrapolation techniques}
\label{sec:extrap}

\subsubsection{Usual extrapolation procedure}

In order to extrapolate the sCI results to the FCI limit, we have adopted the method recently proposed by Holmes, Umrigar and Sharma \cite{Holmes_2017} in the context of the SHCI method. \cite{Holmes_2016, Sharma_2017, Holmes_2017}
It consists of extrapolating the sCI energy, $\EO$, as a function of the second-order Epstein-Nesbet energy, $\EPT$, which is an estimate of the truncation error in the sCI algorithm, i.e $\EPT \approx \EFCI-\EO$. \cite{Huron_1973}
When $\EPT = 0$, the FCI limit has effectively been reached.
This extrapolation procedure has been shown to be robust, even for challenging chemical situations. \cite{Holmes_2017, Sharma_2017, Scemama_2018a, Scemama_2018b, Chien_2018, Garniron_2018, Loos_2018b, Loos_2019} 
Below, we propose an improved extrapolation scheme which \emph{renormalizes} the second-order perturbative correction.

\subsubsection{Renormalized PT2}
\label{sec:rPT2}
At a given sCI iteration, the sCI+PT2 energy is given by
\begin{equation}
	E = \EO + \EPT,
\end{equation}
where $\EO$ and $\EPT$ are given by Eqs.~\eqref{eq:evar} and \eqref{eq:PT2}, respectively.
Let us introduce the following energy-dependent second-order self-energy
\begin{equation}
	\label{eq:E2-w}
	\SigPT[E]
	= \sum_{\alpha} \frac{\Hij{\alpha}{\PsiO}^2}{E - \Hij{\alpha}{\alpha}}.
\end{equation}
Obviously, we have $\SigPT[\EO] = \EPT$.
Now, let us consider the more general problem, which is somewhat related to Brillouin-Wigner perturbation theory, where we have
\begin{equation}
\label{eq:QP}
	E = \EO + \SigPT[E],
\end{equation}
and assume that $\SigPT[E]$ behaves linearly for $E \approx \EO$, i.e.,
\begin{equation}
\label{eq:linearize}
	\SigPT[E] \approx \SigPT[\EO] + (E - \EO) \left. \pdv{\SigPT[E]}{E} \right|_{E = \EO}.
\end{equation}
This linear behavior is corroborated by the findings of Nitzsche and Davidson. \cite{Nitzsche_1978a}
Substituting Eq.~\eqref{eq:linearize} into \eqref{eq:QP} yields
\begin{equation}
\begin{split}
	E & = \EO + \SigPT[\EO] + (E - \EO) \left. \pdv{\SigPT[E]}{E} \right|_{E = \EO}
	\\
	 & = \EO + Z\,\EPT,
\end{split}
\end{equation}
where the renormalization factor is
\begin{equation}
	\label{eq:Z}
	Z = \qty[ 1 - \left. \pdv{\SigPT[E]}{E} \right|_{E = \EO} ]^{-1},
\end{equation}
and
\begin{equation}
	\left. \pdv{\SigPT[E]}{E} \right|_{E = \EO} = - \sum_{\alpha} \frac{\Hij{\alpha}{\PsiO}^2}{(\EO - \Hij{\alpha}{\alpha})^2} < 0.
\end{equation}
Therefore, the renormalization factor fulfills the condition $0 \le Z \le 1$, and its actual computation does not involve any additional cost when computed alongside $\EPT$ as they involve the same quantities.
This renormalization procedure of the second-order correction, that we have named rPT2, bears obvious similarities with the computation of quasiparticle energies within the G$_0$W$_0$ method. \cite{Onida_2002, Reining_2017, Loos_2018a, Veril_2018}
Practically, the effect of rPT2 is to damp the value of $\EPT$ for small wave functions.
Indeed, when $\Ndet$ is small, the sum $\EO + \EPT$ usually overestimates (in magnitude) the FCI energy, yielding a pronounced non-linear behavior of the sCI+PT2 energy.
Consequently, by computing instead the (renormalized) energy $\EO + Z\,\EPT$, one observes a much more linear behavior of the energy, hence an easier extrapolation to the FCI limit.
Its practical usefulness is illustrated in Sec.~\ref{sec:res_extrap}.

\section{Implementation}
\label{sec:implementation}

In this section, we give an overview of the implementation of the various methods present in \QP.
The implementation of the crucial algorithms is explained in detail in the PhD thesis of Dr Y.~Garniron \cite{Garniron_2019} as well as in the Appendix of the present manuscript.

\subsection{Determinant-driven computation of the matrix elements}
\label{sec:detdrive}

For performance sake, it is vital that some basic operations are done efficiently and, notably, the computation of the Hamiltonian matrix elements.
This raises some questions about the data structures chosen to represent the two-electron integrals and determinants, as well as their consequences from an algorithmic point of view.
This section is going to address these questions by going through the basic concepts of our determinant-driven approach.

\subsubsection{Storage of the two-electron integrals}
\label{sec:integrals}

In \QP, the two-electron integrals are kept in memory because they require a fast random access.
Considering the large number of two-electron integrals, a hash table is the natural choice allowing the storage of only non-zero values with a data retrieval in near constant time. \cite{Maurer1975Mar} 
However, standard hashing algorithms tend to shuffle data to limit the probability of collisions.
Here, we favor data locality using the hash function given in Algorithm~\ref{alg:hash}. 
This hash function i) returns the same value for all keys related by permutation symmetry, ii) keeps some locality in the storage of data, and iii) can be evaluated in 10 CPU cycles (estimated with MAQAO\cite{maqao}) if the integer divisions by two are replaced by right bit shift instructions.

\begin{algorithm}
	\caption{Hash function that maps any orbital quartet $(i,j,k,l)$ related by permutation symmetry to a unique integer.}
	\label{alg:hash}
	\SetKwFunction{FMain}{HASH}
	\SetKwProg{Fn}{Function}{:}{}
	\Fn(\tcc*[h]{Hash function for two-electron integrals}){\FMain{$i,j,k,l$}}{
		\KwData{ $i,j,k,l$ are the orbital indices}
		\KwResult{ The corresponding hash}
		$p \gets \min (i,k)$ \;
		$r \gets \max (i,k)$ \;
		$t \gets p + r (r-1)/2$ \;
		$q \gets \min (j,l)$ \;
		$s \gets \max (j,l)$ \;
		$u \gets q + s (s-1)/2$ \;
		$v \gets \min (t,u)$ \;
		$w \gets \max (t,u)$ \;
	\KwRet{$v + w (w-1)/2$} \;
}
\end{algorithm}

The hash table is such that each bucket can potentially store $2^{15}$ consecutive key-value pairs. 
The 15 least significant bits of the hash value are removed to give the bucket index [$i_\text{bucket} = \lfloor \text{hash}(i,j,k,l)/2^{15} \rfloor$], and only those 15 bits need to be stored in the bucket for the key storage [$\text{hash}(i,j,k,l) \mod 2^{16}$].
Hence, the key storage only requires two bytes per key, and they are sorted in increasing order, enabling a binary search within the bucket. 
The key search is always fast since the binary search is bounded by 15 misses and the maximum size of the key array is 64~kiB, the typical size of the L1 cache.
The efficiency of the integral storage is illustrated in Appendix \ref{sec:imple_details_eri}.

\subsection{Internal representation of determinants}
\label{sec:det_representation}

Determinants can be conveniently written as a string of creation operators applied to the vacuum state $\vac$, e.g., $\ac{i} \ac{j} \ac{k} \vac = \kI$.
Because of the fermionic nature of electrons, a permutation of two contiguous creation operators results in a sign change $\ac{j} \ac{i} = -\ac{i} \ac{j}$, which makes their ordering relevant, e.g., $\ac{j} \ac{i} \ac{k} \vac =  -\kI$.
A determinant can be broken down into two pieces of information:
i) a set of creation operators corresponding to the set of occupied spinorbitals in the determinant, and ii) an ordering of the creation operators responsible for the sign of the determinant, known as \emph{phase factor}.
Once an ordering operator $\ordering$ is chosen and applied to all determinants, the phase factor may simply be included in the CI coefficient.

The determinants are built using the following order: i) spin-up ($\uparrow$) spinorbitals are placed before spin-down ($\downarrow$) spinorbitals, as in the Waller-Hartree double determinant representation\cite{Pauncz_1989} $\ordering \kI = \hI \vac = \hI_\uparrow \hI_\downarrow \vac$, and ii) within each operator $\hat{I}_\uparrow$ and $\hat{I}_\downarrow$, the creation operators are sorted by increasing indices.
For instance, let us consider the determinant $\kJ = \ac{j} \ac{k} \ac{\bar i} \ac{i} \vac$ built from the set of spinorbitals $\{i_{\uparrow},j_{\uparrow},k_{\uparrow},i_{\downarrow} \}$ with $i<j<k$.
If we happen to encounter such a determinant, our choice of representation imposes to consider its re-ordered expression $\ordering \kJ = - \ac{i} \ac{j} \ac{k} \ac{\bar i} \vac = -\kJ$, and the phase factor must be handled.

The indices of the creation operators (or equivalently the spinorbital occupations), are stored using the so-called \emph{bitstring} encoding. 
A bitstring is an array of bits; typically, the 64-bit binary representation of an integer is a bitstring of size 64.
Quite simply, the idea is to map each spinorbital to a single bit with value set to its occupation number. 
In other words, 0 and 1 are associated with the \emph{unoccupied} and \emph{occupied} states, respectively.
Additional information about the internal representation of determinants can be found in Appendix \ref{sec:imple_details_det}.

\subsection{
Davidson diagonalization
\label{sec:dav}
}

Finding the lowest root(s) of the Hamiltonian is a necessary step in CI methods.
Standard diagonalization algorithms scale as $\order*{\Ndet^3}$ and $\order*{\Ndet^2}$ in terms of computation and storage, respectively.
Hence, their cost is prohibitive as $\Ndet$ is usually, at least, of the order of few millions.
Fortunately, not all the spectrum of $\hH$ is required: only the first few lowest eigenstates are of interest. 
The Davidson diagonalization \cite{Davidson_1975,Liu_1978,Olsen_1990,Gadea_1994,Crouzeix_1994} is an iterative algorithm which aims at extracting the first $\Nst$ lowest eigenstates of a large matrix. 
This algorithm reduces the cost of both the computation and storage to $\order*{\Nst \Ndet^2}$ and $\order*{\Nst \Ndet}$, respectively.
It is presented as Algorithm \ref{alg:davidson} and further details about the present Davidson algorithm implementation are gathered in Appendix \ref{sec:imple_details_dav}.

\begin{algorithm}
 \caption{Davidson diagonalization procedure.
 Note that $[.,.]$ stands for column-wise matrix concatenation.}
 \label{alg:davidson}
	\SetKwFunction{FMain}{DAVIDSON\_DIAG}
	\SetKwProg{Fn}{Function}{:}{}
\Fn(){\FMain{$\Nst, \mU$}}{
	\KwData{ $\Nst$: Number of requested states}
	\KwData{ $\Ndet$: Number of determinants}
	\KwData{ $\mU$: Guess vectors, $\Ndet \times \Nst$}
	\KwResult{ $\Nst$ lowest eigenvalues eigenvectors of $\mH$ }
\texttt{converged} $\gets \FALSE$ \;
\While{$\neg{\texttt{converged}}$}
{
  Gram-Schmidt orthonormalization of $\mU$ \;
  $\mW \gets \mH\, \mU$ \; 
  $\mh \gets \mU^\dagger\, \mW$ \;
  Diagonalize $\mh$ : eigenvalues $E$ and eigenvectors $\my$ \;
  $\mU' \gets \mU\, \my$ \;
  $\mW' \gets \mW\, \my$ \;
  \For{$k\gets 1,\Nst$}{
    \For{$i\gets 1,\Ndet$} {
      $\mR_{ik}  \gets \frac{E_k \mU_{ik}' - \mW_{ik}'}{\mH_{ii} - E_k}$ \;
    }
  }
  $\texttt{converged} \gets \norm{\mR} < \epsilon$ \;
  $\mU \gets [ \mU, \mR ]$ \;
}
\KwRet{$\mU$}\;
}
\end{algorithm}

\subsection{
CIPSI selection  and PT2 energy
\label{sec:sel}
}

\subsubsection{The basic algorithm}

The largest amount of work for this second version of \QP has been devoted to the improvement of the CIPSI algorithm implementation. \cite{giner:tel-01077016} 
As briefly described in Sec.~\ref{sec:meth}, this is an iterative selection algorithm, where determinants are added to the reference wave function according to a perturbative criterion. 

The $n$th CIPSI iteration can be described as follows:
\begin{enumerate}

	\item The zeroth-order (reference or variational) wave function
	\begin{equation}
		\kPsiO = \sum_{I \in \mcI_n} \cCI{I} \kI
	\end{equation}
                is defined over a set of determinants $\mcI_n$ --- characterized as \emph{internal} determinants --- from which the lowest eigenvector of $\hH$ are obtained.
	
	\item For all \emph{external} determinants $\kalpha \notin \mcI_n$ but connected to $\mcI_n$, i.e., $\mel*{\PsiO}{\hH}{\alpha} \neq 0$, we compute the individual perturbative contribution $\ePT{\alpha}$ given by Eq.~\eqref{eq:e2}.
	This set of external determinants is labeled $\mcA_n$.

	\item Summing the contributions of all the external determinants $\alpha \in \mcA_n$ gives the second-order perturbative correction provided by Eq.~\eqref{eq:PT2} and the FCI energy can be estimated as $\EFCI \approx \EO + \EPT$.

	\item We extract $\kalphas \in \mcAs_n$, the subset of determinants $\kalpha \in \mcA_n$ with the largest contributions $\ePT{\alpha}$, and add them to the variational space $\mcI_{n+1} = \mcI_{n} \cup  \mcAs_n$.
	In practice, in the case of the single-state calculation, we aim at doubling the size of the reference wave function at each iteration.

	\item Iterate until the desired convergence has been reached.

\end{enumerate}

All the details of our current implementation are reported in Appendix \ref{sec:imple_details_sel}.
In the remaining of this section, we only discuss the algorithm of our new stochastic CIPSI selection.

\subsubsection{New stochastic selection}

In the past, CIPSI calculations were only possible in practice thanks to approximations. 
The first approximation restricts the set $\mcA_n$ by defining a set of \emph{generators}.
Indeed, it is very unlikely that $\kalpha$ will be selected if it is not connected to any $\kI$ with a large coefficient, so only the determinants with the largest coefficients are generators.
A second approximation defines a set of \emph{selectors} in order to reduce the cost of $\ePT{\alpha}$ by removing the determinants with the smallest coefficients in the expression of $\PsiO$ in $\EPT$.
This approximate scheme was introduced in the 80's and is known as \emph{three-class CIPSI}.\cite{Evangelisti_1983} 
The downside of these approximations is that the calculation is biased and, consequently, does not strictly converge to the FCI limit. 
Moreover, similar to the initiator approximation in FCIQMC, \cite{Cleland_2010} this scheme suffers from a size-consistency issue. \cite{Tenno_2017} 
The stochastic selection that we describe in this section (asymptotically) cures this problem, as there is \emph{no threshold} on the wave function: if the calculation is run long enough, the unbiased FCI solution is obtained.

Recently, some of us developed a hybrid deterministic/stochastic algorithm for the computation of $\EPT$.\cite{Garniron_2017}
The main idea is to rewrite the expression of
\begin{equation}
	\EPT = \sum_\alpha c_\alpha \mel*{\PsiO}{\hH}{\alpha}
\end{equation}
into elementary contributions labeled by the determinants of the internal space:
\begin{equation}
 \EPT	= \sum_I \sum_{\alpha \in \mcA_I} c_\alpha \mel*{\PsiO}{\hH}{\alpha} 
        = \sum_I \eps_I,
\end{equation}
where
\begin{equation}
        \label{eq:calpha}
	c_\alpha	= \frac{\mel*{\PsiO}{\hH}{\alpha}}{\EO - \Hij{\alpha}{\alpha}}
\end{equation}
is the corresponding coefficient estimated via first-order perturbation theory, and $\mcA_I$ is the subset of determinants $\kalpha$ connected to $\kI$ by $\hH$
such that $\kalpha \notin \cup_{K<I} \mcA_K$.
The sum is decomposed into a stochastic and a deterministic contribution
\begin{equation}
        \label{eq:edetstoch}
	\EPT = \sum_{J \in \mcD} \eps_J + \sum_{K \in \mcS} \eps_K,
\end{equation}
where $\mcD$ and $\mcS$ are the sets of determinants included in the deterministic and stochastic components, respectively.

The $\kI$'s are sorted by decreasing $\cCI{I}^2$, and two processes are used simultaneously to compute the contributions $\eps_I$.
The first process is stochastic and $\kI$ is drawn according to $\cCI{I}^2$.
When a given $\eps_I$ has been computed once, its contribution is stored such that if $\kI$ is drawn again later the contribution does not need to be recomputed.
The only update is to increment the number of times it has been drawn for the Monte Carlo statistics.
In parallel, a deterministic process is run, forcing to compute the contribution $\eps_I$ with the smallest index which has yet to be computed.
The deterministic component is chosen as the first contiguous set of $\eps_I$.
Hence, the computation of $\EPT$ is unbiased, and the exact deterministic value can be obtained in a finite time if the calculation is run long enough. 
The stochastic part is only a convergence accelerator providing a reliable error bar.
The computation of $\EPT$ is run with a default stopping criterion set to $\abs*{\delta \EPT / \EPT}=0.002$, where $\delta \EPT$ is the statistical error associated with $\EPT$.
We would like to stress that, thanks to the present semistochastic algorithm, the complete wave function is considered, and that no threshold is required.
Consequently, size-consistency will be preserved if a size-consistent perturbation theory is applied.

While performing production runs, we have noticed that the computation of $\EPT$ was faster than the CIPSI selection.
Hence, we have slightly modified the routines computing $\EPT$ such that the selection of determinants is performed alongside the computation of $\EPT$.
This new on-the-fly CIPSI selection performed during the stochastic PT2 calculation completely removes the conventional (deterministic) selection step, and the determinants are selected with no additional cost.
We have observed that, numerically, the curves of the variational energy as a function of $\Ndet$ obtained with either the deterministic or the stochastic selections are indistinguishable, so that the stochastic algorithm does not harm the selection's quality.

For the selection of multiple states, one PT2 calculation is run for each state and, as proposed by Angeli \textit{et al}., \cite{Angeli_1997} the selection criterion is modified as
\begin{equation}
	\tePT{\alpha} = \sum_{k}^{\Nst} \frac{c_{\alpha k}}{\max_I{c_{I k}^2}} \mel*{\PsiO_{k}}{\hH}{\alpha},
\end{equation}
with
\begin{equation}
	c_{\alpha k} = \frac{\mel*{\PsiO_k}{\hH}{\alpha}}{\mel*{\PsiO_k}{\hH}{\PsiO_k} - \Hij{\alpha}{\alpha}}.
\end{equation}
This choice gives a balanced selection between states of different multi-configurational nature.

\section{spin-adapted wave functions}
\label{sec:cipsi_s2}

Determinant-based sCI algorithms generate wave functions expressed in a truncated space of determinants. 
Obviously, the selection presented in the previous section does not enforce that $\hH$ commutes with $\hS^2$ in the truncated space.
Hence, the eigenstates of $\hH$ are usually not eigenvectors of $\hS^2$, although the situation improves when the size of the internal space tends to be complete.
A natural way to circumvent this problem is to work in the basis of \emph{configuration state functions} (CSFs), but this representation makes the direct computation of the Hamiltonian less straightforward during the Davidson diagonalization.

Instead, we follow the same path as the MELD and SCIEL codes,\cite{Davidson_1979,Kozlowski_1994,Caballol_1998} and identify all the spatial occupation patterns associated with the determinants.\cite{Applencourt_2019} 
We then generate all associated spin-flipped configurations, and add to the internal space all the missing determinants.
This procedure ensures that $\hH$ commutes with $\hS^2$ in the truncated space, and spin-adapted states are obtained by the diagonalization of $\hH$.
In addition, we apply a penalty method in the diagonalization by modifying the Hamiltonian as \cite{Fales2017Sep} 
\begin{equation}
	\tilde{\mH} = \mH + \gamma \qty( \mS^2 - \mI \langle S^2 \rangle_\text{target} )^2,
\end{equation}
where $\mI$ is the identity matrix and $\gamma$ is a fixed parameter set to 0.1 by default.
This improves the convergence to the desired spin state, but also separates degenerate states with different spins, a situation that can potentially occurs with Rydberg states.
In the Davidson algorithm, this requires the additional computation of $\mS^2\, \mU$, for which the cost is expected to be the same as the cost of $\mH \, \mU$ (see Algorithm \ref{alg:davidson}).
The cost of computing $\mH \, \mU$ and $\mS^2\, \mU$ is mostly due to the search of the connected pairs of determinants, namely the determinants $\bI$ and $\kJ$ for which $\Hij{I}{J}$ and $\Sij{I}{J}$ are not zero due to Slater-Condon's rules.
We have modified the function computing $\mH \, \mU$ so that it also computes $\mS^2\, \mU$ at the same time. Hence, the search of connected pairs is done once for both operations and $\mS^2\, \mU$ is obtained with no extra computational cost.

Working with spin-adapted wave functions increases the size of the internal space by a factor usually between 2 and 3, but it is particularly important if one is willing to obtain excited states. \cite{Loos_2018b, Scemama_2018a, Scemama_2018b, Loos_2019} Therefore, the default in {\QP} is to use spin-adapted wave functions.

\section{Parallelism}
\label{sec:parallelism}

In \QP, multiple parallelism layers are implemented: a fine-grained layer to benefit from shared memory, an intermediate layer to benefit from fast communication within a group of nodes, and a coarse-grained layer to interconnect multiple groups of nodes.
Fine-grained parallelism is performed with OpenMP\cite{openmp} in almost every single routine.
Then, to go beyond a single compute node,
\QP does not use the usual single {program/multiple} data (SPMD) paradigm.
A task-based parallelism framework is implemented with the ZeroMQ library.\cite{Hintjens_2013}
The single-node instance runs a compute process as well as a task server process, while helper programs can be spawned asynchronously on different (heterogeneous) machines to run a distributed calculation.
The helper programs can connect via ZeroMQ to the task server at any time, and contribute to a running calculation.
As the ZeroMQ library does not take full advantage of the low latency hardware present in HPC facilities, the helper programs are parallelized also with the message passing interface\cite{MPI} (MPI) for fast communication among multiple client nodes, typically for fast broadcasting of large data structures.

Hence, we have 3 layers of parallelism in {\QP}: OpenMP, MPI and ZeroMQ.
This allows for an elastic management of resources: a running calculation taking too much time can be dynamically accelerated by plugging in more computing resources, by submitting more jobs in the queue or possibly in the cloud, i.e.~outside of the HPC facility.
This scheme has the advantage that it is not necessary to wait for all the nodes to be free to start a calculation, and hence minimizes the waiting time in the batch queue.
It also gives the possibility to use altogether different helper programs.
For instance, one could use a specific GPU-accelerated helper program on a GPU node while CPU-only helpers run on the CPU-only partition of the cluster.
It is also possible to write a helper program that helps only one PT2/selection step and then exit, allowing to gather resources after the PT2/selection has started, and freeing them for the following diagonalization step.

The current limitation of {\QP} is the memory of the single-node instance. 
We have not yet considered the possibility to add more compute nodes to increase the available memory, but this can be done by transforming the main program into an MPI program using scattered data structures.

We now describe how the Davidson and PT2/selection steps are parallelized.

\subsubsection{Davidson diagonalization}

In the direct Davidson diagonalization method, the computational bottleneck is the matrix product $\mW = \mH\,\mU$, and only this step needs to be distributed.
The calculation is divided into independent tasks where each task builds a unique piece of $\mW$ containing 40\,000 consecutive determinants. 
Communicating the result of all the tasks scales as $\order*{\Ndet}$, independently of the number of parallel processes.
On the other hand, $\mU$ needs to be broadcast efficiently at the beginning of the calculation to each slave process.

The computation of a task is parallelized with OpenMP, looping in a way that guarantees a safe write access to $\mW$, avoiding the need of a lock.
When idle, a slave process requests a task to the ZeroMQ task server, computes the corresponding result and sends it to the \emph{collector} thread of the master instance via ZeroMQ.
As the OpenMP tasks are not guaranteed to be balanced, we have used a dynamic scheduling, with a chunk size of 64 elements. 
The reason for this chunk size is to force that multiple threads access to $\mW$ at memory addresses far apart, avoiding the so-called \emph{false sharing} performance degradation that occurs when multiple threads write simultaneously in the same cache line. \cite{Bolosky1993Sep} 
When the task is fully computed, the computed piece of $\mW$ is sent back to the master process and a new task is requested, until the task queue is empty.

\begin{figure}
	\begin{center}
		\includegraphics[width=\linewidth]{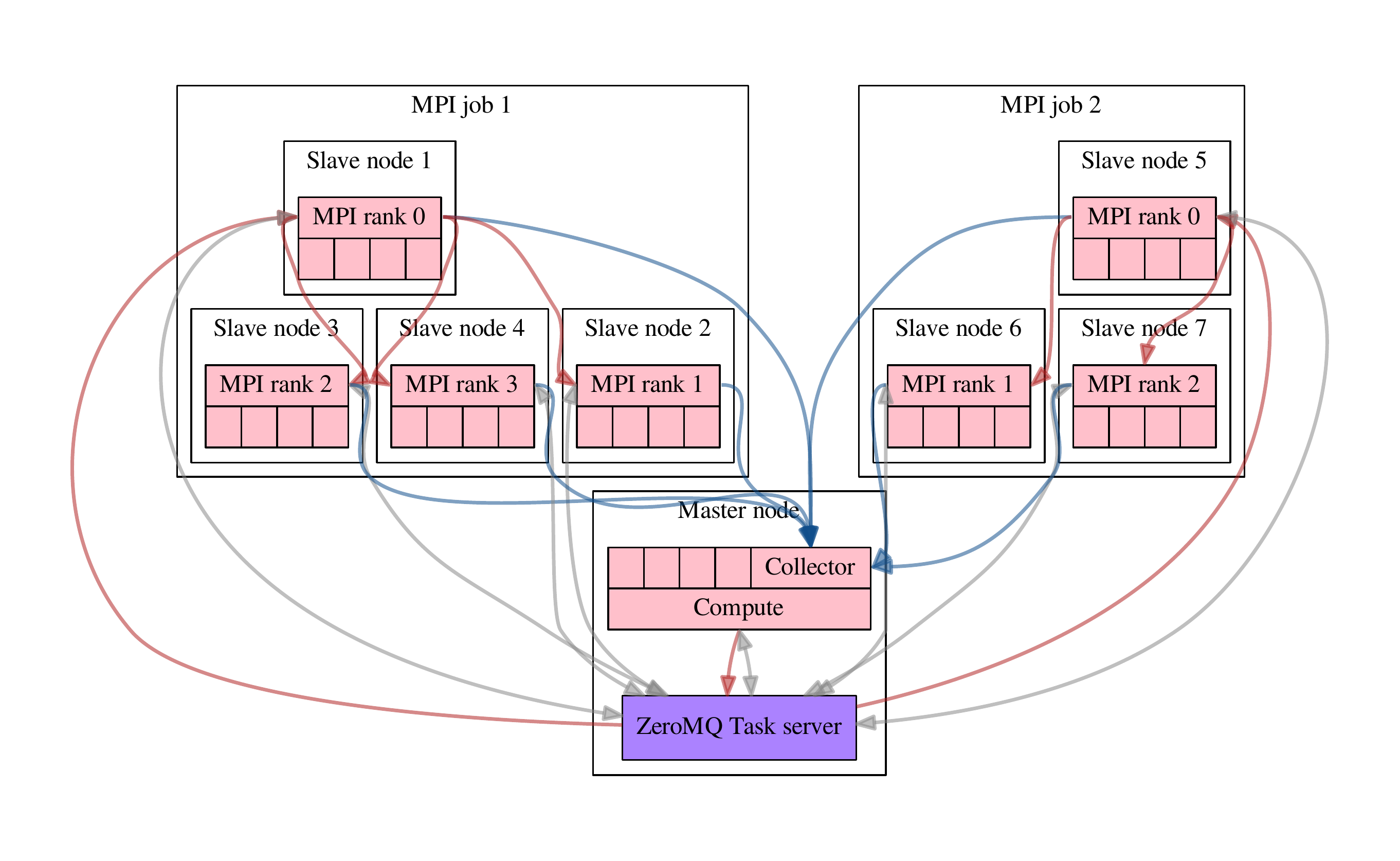}
		\caption{Communications in the Davidson diagonalization for a calculation with a master node and two helper MPI jobs, each using 4 cores for the computation.
Red arrows represent the broadcast of $\mU$ starting from the compute process of the master node, gray arrows the exchange of ZeroMQ messages with the task server and blue arrows the collection of the results.}
		\label{fig:davidson_comm}
	\end{center}
\end{figure}

The $\mU$ and $\mW$ arrays are shared among threads, as well as all the large constant data needed for the calculation such as the ERIs.
Sharing $\mU$ also provides the benefit to reduce the amount of communication since $\mU$ needs to be fetched only once for each node, independently of the number of cores.
To make the broadcast of $\mU$ efficient, the slave helper program is parallelized with MPI in a SPMD fashion, and each node runs a single MPI process.
The $\mU$ matrix is fetched from the ZeroMQ server by the process with rank zero, and then it is broadcast to the other slave processes within the same MPI job via MPI primitives.
Then, each MPI process behaves independently and communicates via ZeroMQ with the task server, and with the master node which collects the results.
A schematic view of the communication is presented in Fig~\ref{fig:davidson_comm}.

\subsubsection{CIPSI selection and PT2 energy}
\label{seq:fragmentation}

In the computation of $\EPT$ and the CIPSI selection, each task corresponds to the computation of one $\eps_J$ or $\eps_K$ in Eq.~\eqref{eq:edetstoch}, together with the selection of the associated external determinants.  
To establish the list of tasks, the Monte Carlo sampling is pre-computed on the master node.
We associate to each task the number of drawn Monte Carlo samples such that running averages can be computed when the results of the tasks have been received by the collector thread.
When the convergence criterion is reached, the task queue is emptied and the collector waits for all the running tasks to terminate.

\begin{figure}
	\begin{center}
		\includegraphics[width=\linewidth]{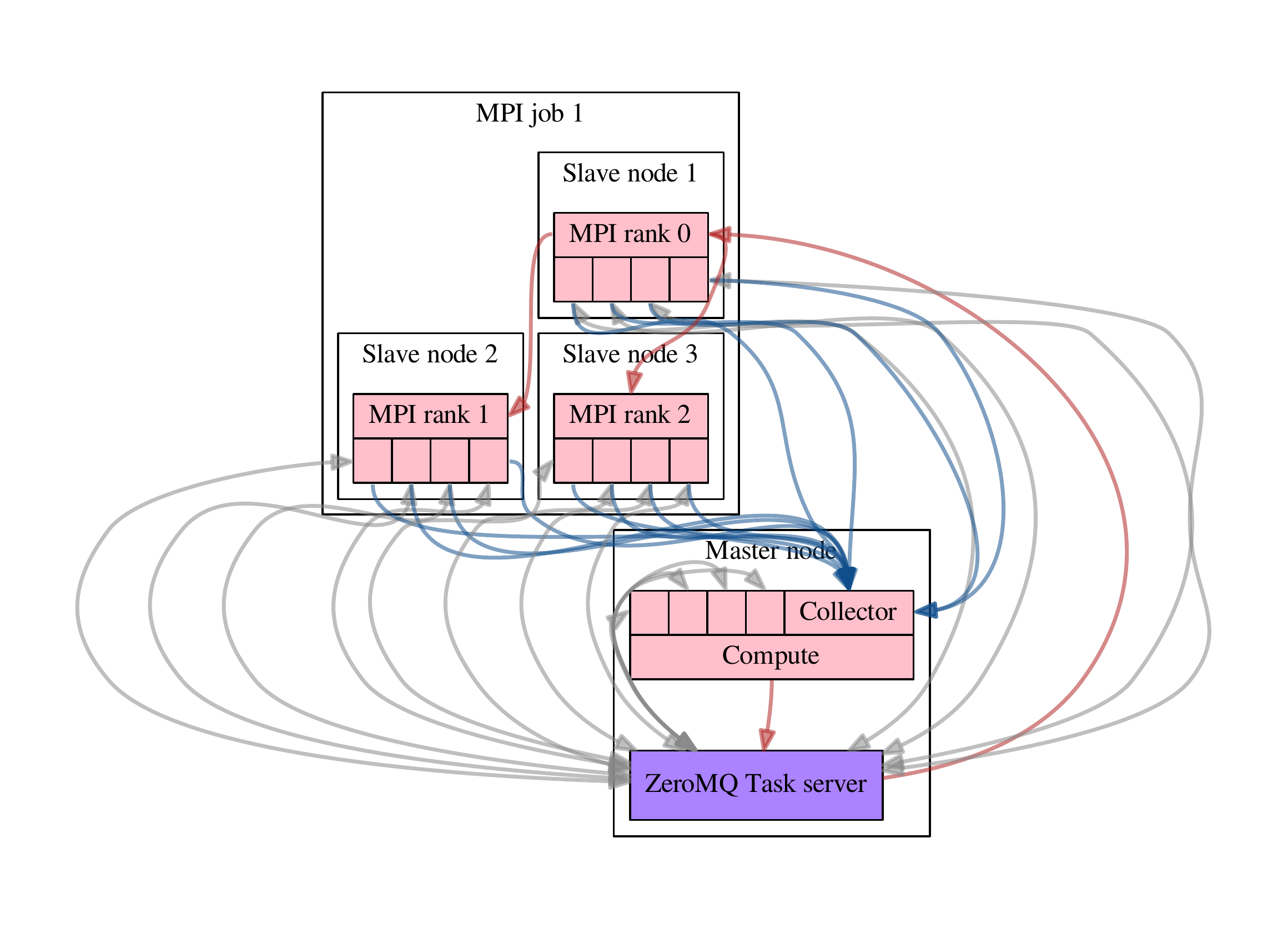}
		\caption{Communications in the stochastic selection for a calculation with a master node and one helper MPI job, each using 4 cores for the computation.
Red arrows represent the broadcast of the common data starting from the compute process of the master node, gray arrows the exchange of ZeroMQ messages with the task server and blue arrows the collection of the results.}
		\label{fig:pt2_comm}
	\end{center}
\end{figure}

As opposed to the Davidson implementation where each task is parallelized with OpenMP, here each OpenMP thread handles independently a task computed on a single core.
Hence, there are multiple ZeroMQ clients per node, typically one per core, requesting tasks to the task server and sending the results back to the collector thread (see Fig.~\ref{fig:pt2_comm}).
Here, all the OpenMP threads are completely independent during the whole selection, and this explains the pleasing scaling properties of our implementation, as shown in Sec.~\ref{sec:perf}.
As in the Davidson distributed scheme, when the helper programs are run with MPI all the common data are communicated once from the ZeroMQ server to the rank-zero MPI process.
Then, the data is broadcast to all the other processes with MPI primitives (there is one MPI process per node).

\section{Results}

\subsection{Capabilities of \QP}
\label{sec:cap_QP}
Before illustrating the new features of \QP in the next subsection.
We propose to give an overview of what can be achieved (in terms of system and basis set sizes) with the current implementation of \QP.
To do so we propose to review some of our very recent studies.

In Ref.~\onlinecite{Loos_2018b}, we studied 18 small molecules (water, hydrogen sulfide, ammonia, hydrogen chloride,
dinitrogen, carbon monoxide, acetylene, ethylene, formaldehyde, methanimine, thioformaldehyde, acetaldehyde, cyclopropene, diazomethane, formamide, ketene, nitrosomethane, and the smallest streptocyanine) with sizes ranging from 1 to 3 non-hydrogen atoms.
For such systems, using sCI expansions of several million determinants, we were able to compute more than hundred highly accurate vertical excitation energies with typically augmented triple-$\zeta$ basis sets.
It allowed us to benchmark a series of 12 state-of-the-art excited-state wave function methods accounting for double and triple excitations.

Even more recently, \cite{Loos_2019} we provided accurate reference excitation energies for transitions involving a
substantial amount of double excitation using a series of increasingly large diffuse-containing atomic basis sets. 
Our set gathered 20 vertical transitions from 14 small- and medium-size molecules (acrolein, benzene, beryllium atom, butadiene, carbon dimer and trimer, ethylene, formaldehyde, glyoxal, hexatriene, nitrosomethane, nitroxyl, pyrazine, and tetrazine).
For the smallest molecules, we were able to obtain well converged excitation energies with augmented quadruple-$\zeta$ basis set while only augmented double-$\zeta$ bases
were manageable for the largest systems (such as acrolein, butadiene, hexatriene and benzene).
Note that the largest sCI expansion considered in this study had more than 200 million determinants.

In Ref.~\onlinecite{giner_cu_2018}, Giner \textit{et al.}~studied even larger systems containing transition metals: \ce{[CuCl4]^2-}, \ce{[Cu(NH3)4]^2+} and  \ce{[Cu(H2O)4]^2+}. 
They were able, using large sCI expansions, to understand the physical phenomena that determine the relative energies of three of the lowest electronic states of each of these square-planar copper complexes.

\subsection{Extrapolation}
\label{sec:res_extrap}

To illustrate the extrapolation procedure described in Sec.~\ref{sec:extrap}, we consider a cyanine dye \cite{LeGuennic_2015} \ce{H2N-CH=NH2+} (labeled as CN3 in the remaining) in both its ground state and first excited state. \cite{Send_2011, Boulanger_2014, Garniron_2018}
The geometry is the equilibrium geometry of the ground state optimized at the PBE0/cc-pVQZ level. \cite{Boulanger_2014}
The ground state is a closed shell, well described by a single reference, while the excited state is singly excited and requires, at least, two determinants to be properly modeled.  
The calculations were performed in the aug-cc-pVDZ basis set with state-averaged natural orbitals obtained from an initial CIPSI calculation.
All the electrons were correlated, so the FCI space which is explored corresponds to a CAS(24,114) space. 
The reference excitation energy, obtained at the CC3/ANO-L-VQZP level is 7.18~eV \cite{Send_2011} (see also Ref.~\onlinecite{Garniron_2018}).
Note that this particular transition is fairly insensitive to the basis set as long as at least one set of diffuse functions is included.
For example, we have obtained 7.14 and 7.13~eV at the CC3/aug-cc-pVDZ and CC3/aug-cc-pVTZ levels, respectively. \cite{Loos_2018b}

\begin{figure*}
	\begin{minipage}{0.6\linewidth}
		\includegraphics[width=\linewidth]{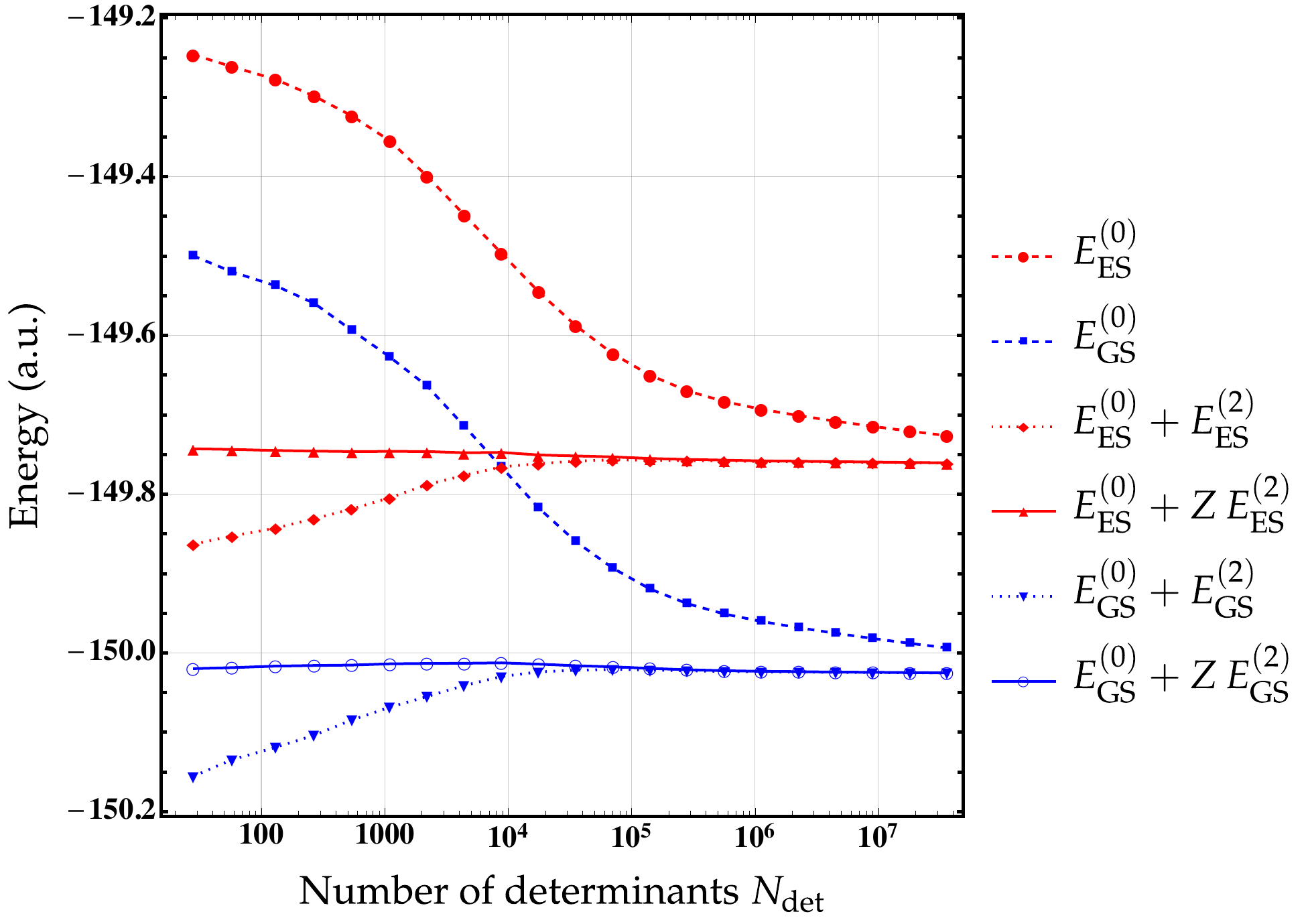}
	\end{minipage}
	\begin{minipage}{0.35\linewidth}
		\includegraphics[width=\linewidth]{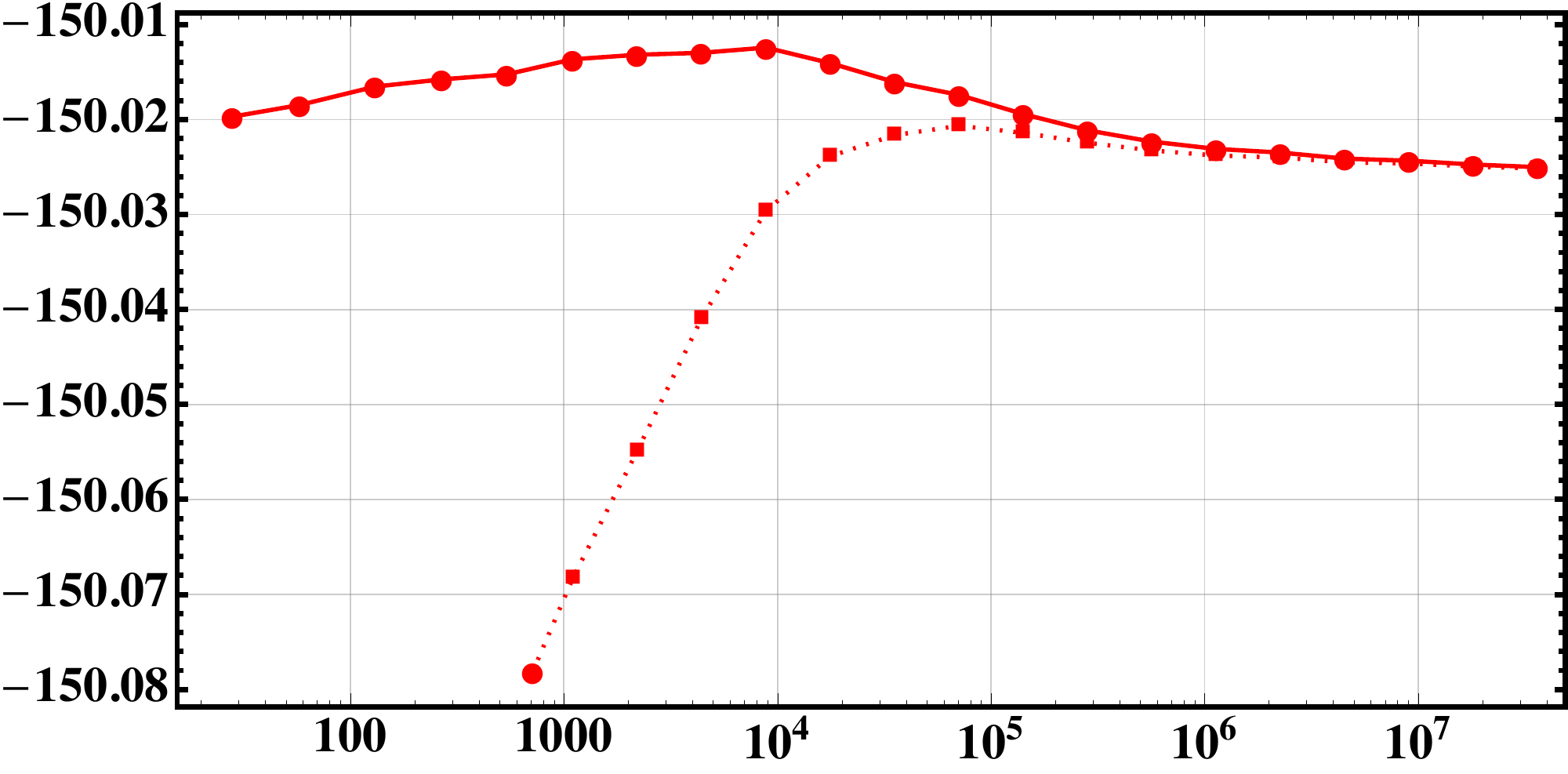}
		\\
		\includegraphics[width=\linewidth]{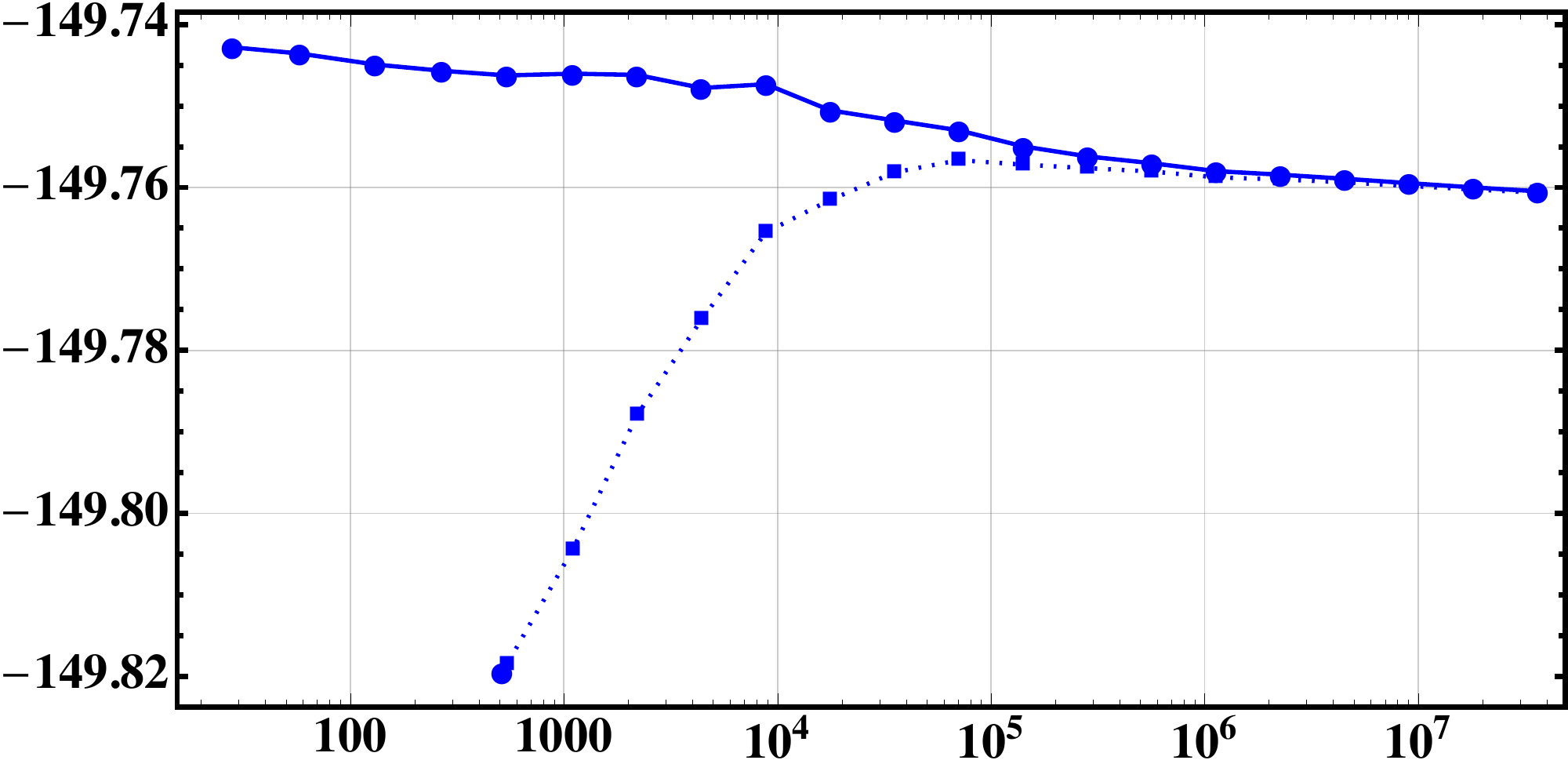}
	\end{minipage}
		\caption{Energy convergence of the ground state (GS, in blue) and excited state (ES, in red) of CN3 with respect to the number of determinants $\Ndet$ in the reference space.
		The zeroth-order energy $\EO$ (dashed) , its second-order corrected value $\EO+\EPT$ (dotted) as well as its renormalized version $\EO+Z \EPT$ (solid) are represented.
		See Table \ref{tab:energy_pt2} for raw data.}
		\label{fig:energy_pt2}
\end{figure*}

\begin{figure*}
	\begin{minipage}{0.45\linewidth}
		\includegraphics[width=\linewidth]{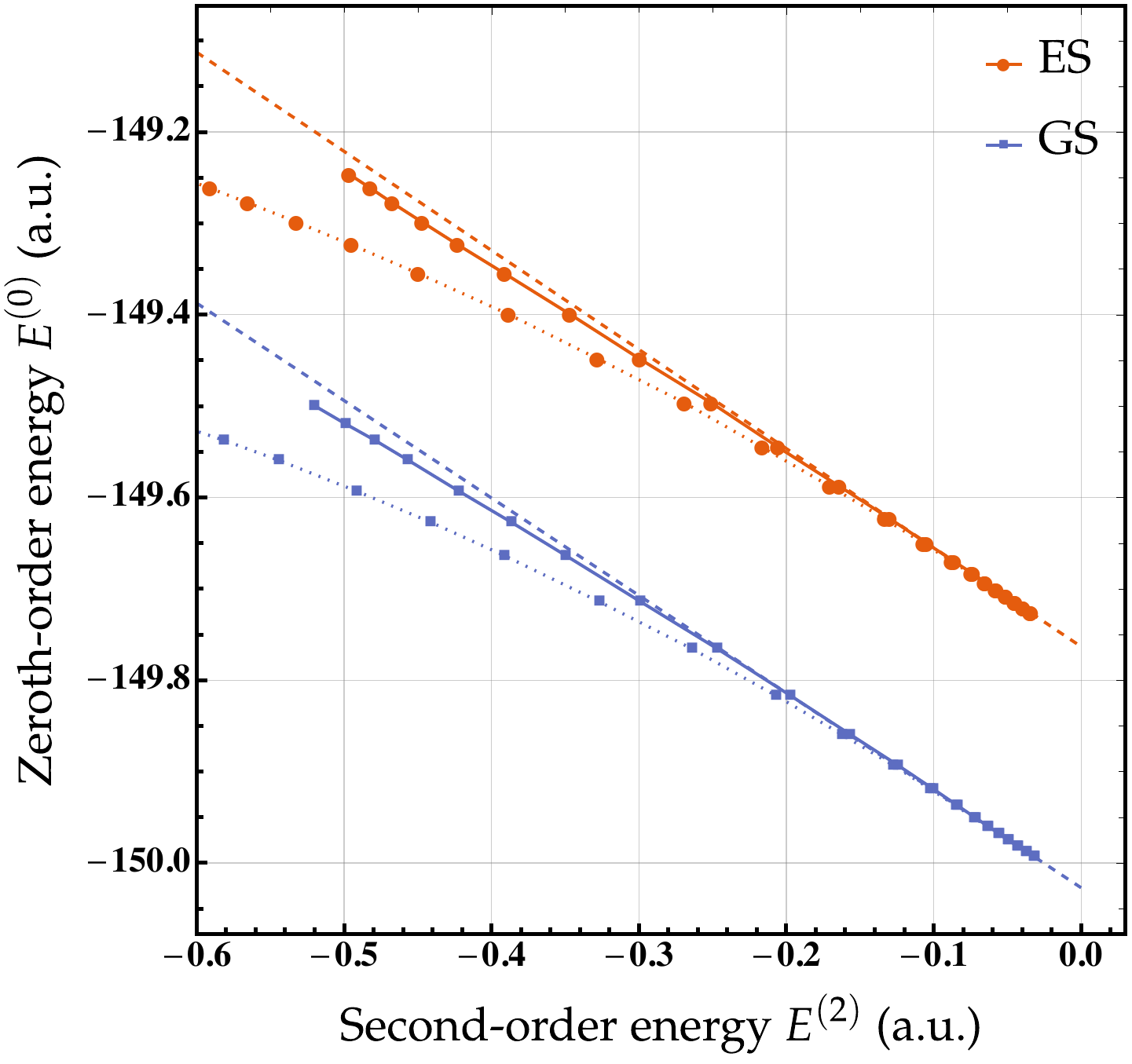}
	\end{minipage}
	\begin{minipage}{0.45\linewidth}
		\includegraphics[width=\linewidth]{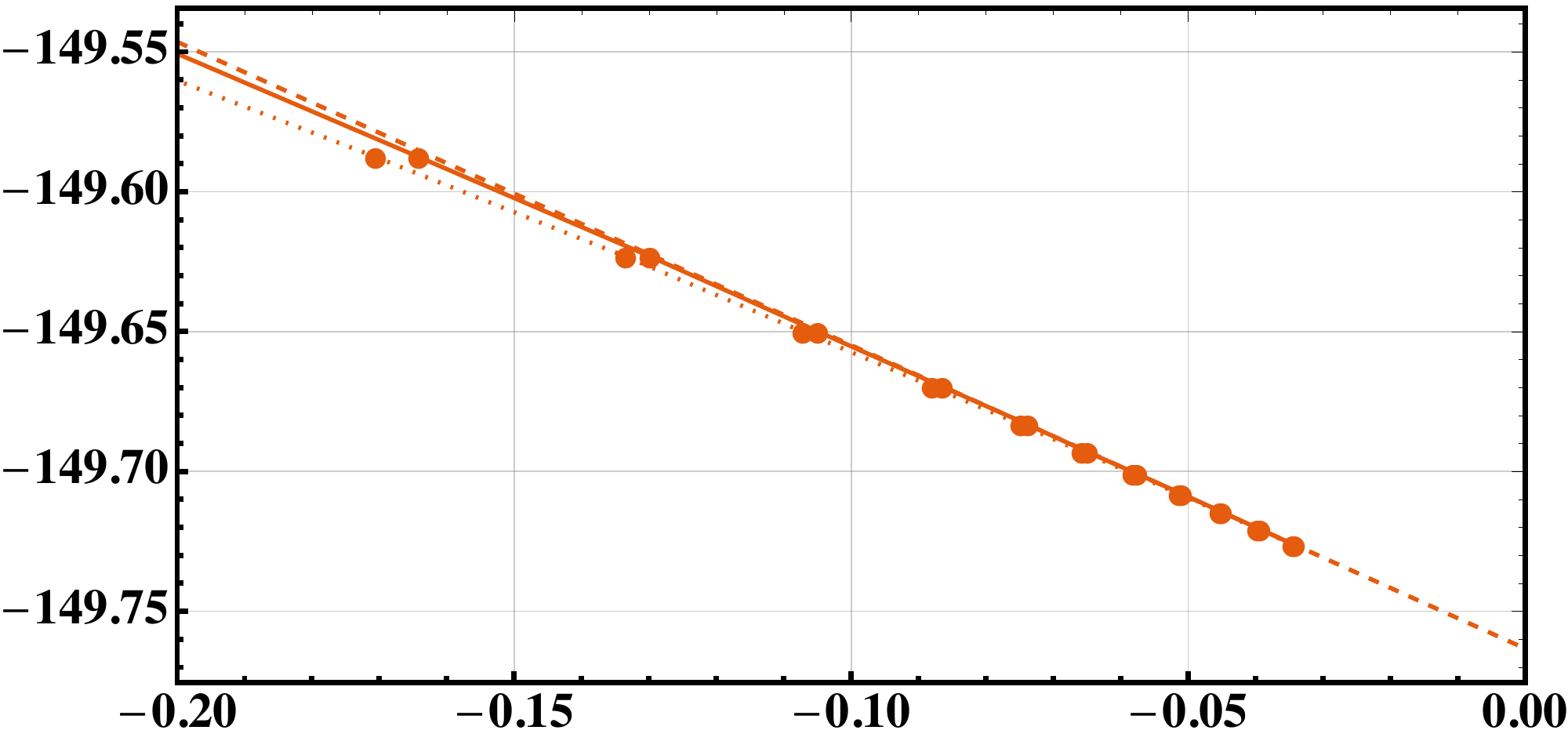}
		\\
		\includegraphics[width=\linewidth]{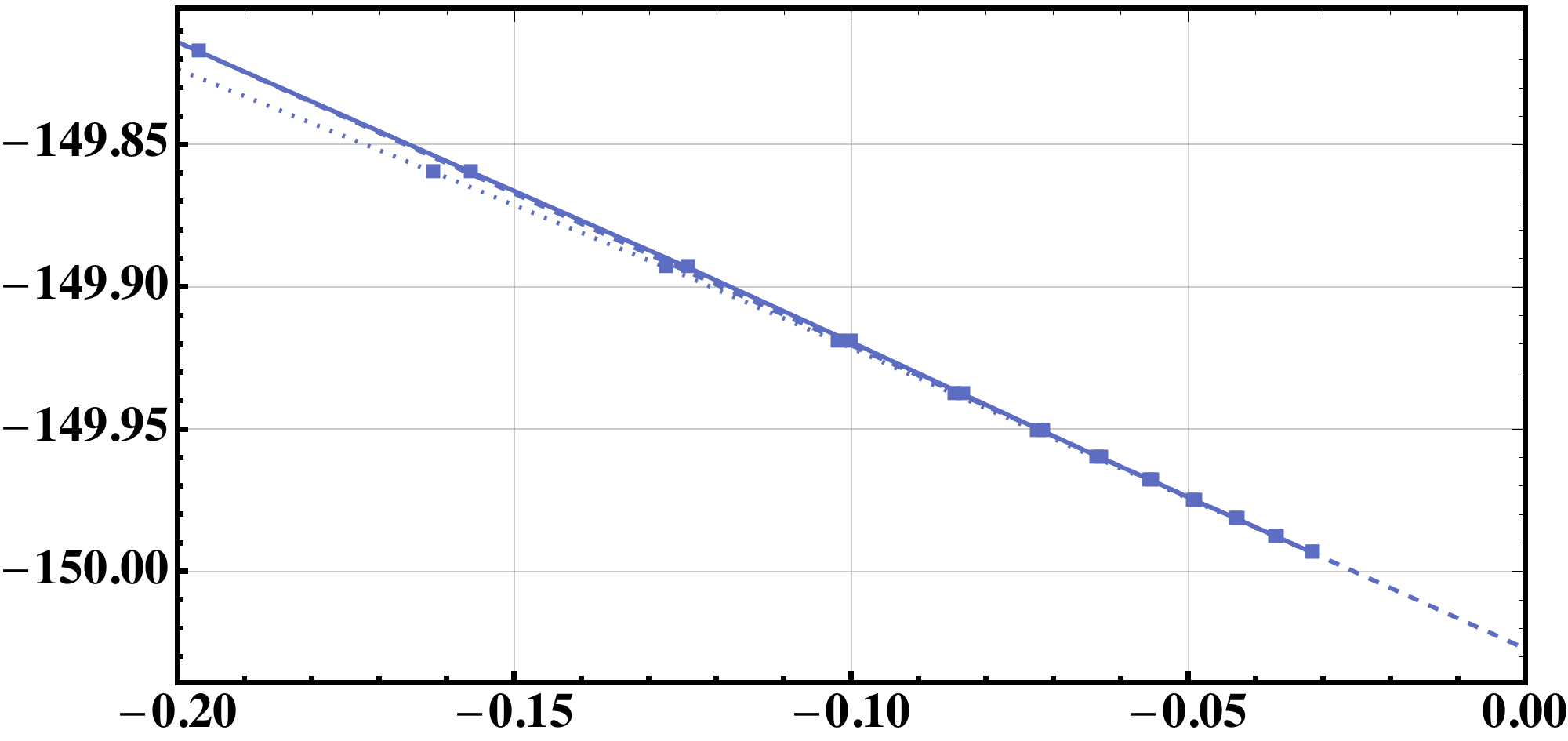}
	\end{minipage}
	\caption{Zeroth-order energy $\EO$ as a function of the second-order energy $\EPT$ (dotted lines) or its renormalization variant $Z\,\EPT$ (solid lines).
	A linear fit (dashed lines) of the last 6 points is also reported for comparison.
	See Table \ref{tab:energy_pt2} for raw data.}
	\label{fig:extrap}
\end{figure*}

In Fig.~\ref{fig:energy_pt2}, we plot the energy convergence of the ground state (GS) and the excited state (ES) as a function of the number of determinants $\Ndet$, with and without the second-order perturbative contribution.
From the data gathered in Table \ref{tab:energy_pt2}, one can see that, although $\EPT$ is still large (roughly \InAU{0.02}), the sCI+PT2 and sCI+rPT2 excitation energies converge to a value of 7.20~eV compatible with the reference energy obtained in a larger basis set.
We have also plotted the sCI+rPT2 energy given by $\EO + Z \EPT$ (see Sec.~\ref{sec:rPT2}) and we clearly see that this quantity converges much faster than the usual sCI+PT2 energy.
Even for very small reference wave function, the energy gap between GS and ES is qualitatively correct.
The graph of Fig.~\ref{fig:extrap}, which shows the zeroth-order energy $\EO$ as a function of the second-order energy $\EPT$ (dotted lines) or its renormalization variant $Z\,\EPT$ (solid lines), also indicates that it is practically much easier to extrapolate to the FCI limit using the rPT2 correction.

\begin{squeezetable}
\begin{table*}
\caption{Zeroth-order energy $\EO$, second-order perturbative correction $\EPT$ and its renormalized version $Z \EPT$ (in hartree) of CN3 for increasingly large wave functions. 
The excitation energy $\Delta E$ (in eV) is the energy difference between the ground state (GS) and the excited state (ES).
The statistical error, corresponding to one standard deviation, is reported in parenthesis.}
\label{tab:energy_pt2}
\begin{ruledtabular}
\begin{tabular}{rdddddddd}
				&	\multicolumn{2}{c}{$\EO$}  
				&	\multicolumn{3}{c}{$\EO + \EPT$}
				&	\multicolumn{3}{c}{$\EO + Z \EPT$}\\
				\cline{2-3}	\cline{4-6} \cline{7-9}
\tabc{$\Ndet$}	& \tabc{GS (a.u.)} & \tabc{ES (a.u.)}  
			 	& \tabc{GS (a.u.)} & \tabc{ES (a.u.)} & \tabc{$\Delta E$ (eV)} 
			 	& \tabc{GS (a.u.)} & \tabc{ES (a.u.)} & \tabc{$\Delta E$ (eV)} \\
\hline
 $28$ 			& -149.499\,574 & -149.246\,268 & -150.155(1)     & -149.863(1)     & 7.95(5)  & -150.020(1)     & -149.743(1)     & 7.54(5)  \\
 $58$ 			& -149.519\,908 & -149.261\,390 & -150.134(1)     & -149.853(1)     & 7.67(5)  & -150.018(1)     & -149.744(1)     & 7.48(5)  \\
 $131$ 			& -149.537\,424 & -149.277\,496 & -150.118(1)     & -149.842\,7(9)  & 7.52(4)  & -150.017(1)     & -149.744\,9(9)  & 7.39(4)  \\
 $268$ 			& -149.559\,465 & -149.298\,484 & -150.103\,5(9)  & -149.830\,8(9)  & 7.42(4)  & -150.015\,8(9)  & -149.745\,7(9)  & 7.35(4)  \\
 $541$ 			& -149.593\,434 & -149.323\,302 & -150.084\,5(8)  & -149.818\,6(8)  & 7.24(4)  & -150.015\,2(8)  & -149.746\,3(8)  & 7.32(4)  \\
 $1\,101$ 		& -149.627\,202 & -149.354\,807 & -150.068\,3(8)  & -149.804\,5(8)  & 7.18(3)  & -150.013\,7(8)  & -149.746\,0(8)  & 7.28(3)  \\
 $2\,207$ 		& -149.663\,850 & -149.399\,522 & -150.054\,9(7)  & -149.787\,9(7)  & 7.26(3)  & -150.013\,2(7)  & -149.746\,2(7)  & 7.27(3)  \\
 $4\,417$ 		& -149.714\,222 & -149.448\,133 & -150.040\,9(6)  & -149.776\,2(6)  & 7.20(3)  & -150.013\,0(6)  & -149.747\,8(6)  & 7.22(3)  \\
 $8\,838$ 		& -149.765\,886 & -149.496\,401 & -150.029\,6(5)  & -149.765\,5(5)  & 7.19(2)  & -150.012\,4(5)  & -149.747\,3(5)  & 7.21(2)  \\
 $17\,680$ 		& -149.817\,301 & -149.545\,048 & -150.023\,9(4)  & -149.761\,5(4)  & 7.14(2)  & -150.014\,1(4)  & -149.750\,5(4)  & 7.17(2)  \\
 $35\,380$ 		& -149.859\,737 & -149.587\,668 & -150.021\,6(3)  & -149.758\,2(3)  & 7.17(1)  & -150.016\,1(3)  & -149.751\,8(3)  & 7.19(1)  \\
 $70\,764$ 		& -149.893\,273 & -149.623\,235 & -150.020\,7(2)  & -149.756\,6(3)  & 7.18(1)  & -150.017\,4(2)  & -149.753\,0(3)  & 7.19(1)  \\
 $141\,545$ 		& -149.919\,463 & -149.650\,109 & -150.021\,4(2)  & -149.757\,2(2)  & 7.189(8) & -150.019\,4(2)  & -149.755\,0(2)  & 7.196(8) \\
 $283\,108$ 		& -149.937\,839 & -149.669\,735 & -150.022\,4(2)  & -149.757\,6(2)  & 7.206(7) & -150.021\,1(2)  & -149.756\,2(2)  & 7.209(7) \\
 $566\,226$ 		& -149.950\,918 & -149.683\,278 & -150.023\,3(1)  & -149.758\,0(1)  & 7.217(6) & -150.022\,3(1)  & -149.757\,0(1)  & 7.219(6) \\
 $1\,132\,520$ 		& -149.960\,276 & -149.693\,053 & -150.023\,8(1)  & -149.758\,8(1)  & 7.212(5) & -150.023\,1(1)  & -149.758\,0(1)  & 7.214(5) \\
 $2\,264\,948$ 		& -149.968\,203 & -149.700\,907 & -150.024\,0(1)  & -149.759\,0(1)  & 7.211(4) & -150.023\,5(1)  & -149.758\,4(1)  & 7.212(4) \\
 $4\,529\,574$ 		& -149.975\,230 & -149.708\,061 & -150.024\,5(1)  & -149.759\,4(1)  & 7.215(4) & -150.024\,1(1)  & -149.758\,9(1)  & 7.216(4) \\
 $9\,057\,914$ 		& -149.981\,770 & -149.714\,526 & -150.024\,63(9) & -149.759\,81(8) & 7.206(3) & -150.024\,34(9) & -149.759\,48(8) & 7.207(3) \\
 $18\,110\,742$		& -149.987\,928 & -149.720\,648 & -150.024\,95(7) & -149.760\,25(8) & 7.203(3) & -150.024\,74(7) & -149.760\,00(8) & 7.204(3) \\
 $36\,146\,730$		& -149.993\,593 & -149.726\,253 & -150.025\,27(6) & -149.760\,65(7) & 7.198(3) & -150.025\,02(6) & -149.760\,47(7) & 7.198(3) \\
\end{tabular}
\end{ruledtabular}
\end{table*}
\end{squeezetable}

\begin{squeezetable}
\begin{table}
\caption{Zeroth-order energy $\EO$, second-order perturbative correction $\EPT$ and its renormalized version $Z \EPT$ (in hartree) as a function of the number of determinants $\Ndet$ for the ground-state of the chromium dimer \ce{Cr2} computed in the cc-pVQZ basis set. 
The statistical error, corresponding to one standard deviation, is reported in parenthesis.}
\label{tab:Cr2}
\begin{ruledtabular}
\begin{tabular}{rddd}
\tabc{$\Ndet$}	&	\tabc{$\EO$}	&	\tabc{$\EO + \EPT$}	&	\tabc{$\EO + Z \EPT$}		
				\\
\hline
     $1\,631$       & -2086.742\,321    &       -2087.853(3)      &            -2087.679(2)     \\
     $3\,312$       & -2086.828\,496    &       -2087.821(2)      &            -2087.688(1)     \\
     $6\,630$       & -2086.920\,161    &       -2087.792(1)      &            -2087.694(1)     \\
    $13\,261$       & -2087.008\,701    &       -2087.764(1)      &            -2087.694(1)     \\
    $26\,562$       & -2087.091\,669    &       -2087.743(1)      &            -2087.692(1)     \\
    $53\,129$       & -2087.165\,533    &       -2087.725(1)      &            -2087.689(1)     \\
   $106\,262$       & -2087.234\,564    &       -2087.710\,2(9)   &            -2087.685\,0(8)  \\
   $212\,571$       & -2087.293\,488    &       -2087.703\,0(8)   &            -2087.685\,0(7)  \\
   $425\,185$       & -2087.343\,762    &       -2087.697\,3(7)   &            -2087.684\,4(7)  \\
   $850\,375$       & -2087.386\,276    &       -2087.697\,8(6)   &            -2087.688\,1(6)  \\
$1\,700\,759$       & -2087.422\,707    &       -2087.698\,9(6)   &            -2087.691\,6(5)  \\
$3\,401\,504$       & -2087.454\,427    &       -2087.700\,7(5)   &            -2087.695\,1(5)  \\
$6\,802\,953$       & -2087.482\,238    &       -2087.703\,2(4)   &            -2087.698\,8(4)  \\
$13\,605\,580$      & -2087.506\,838    &       -2087.705\,6(4)   &            -2087.702\,2(4)  \\
$27\,210\,163$      & -2087.528\,987    &       -2087.709\,2(4)   &            -2087.706\,4(4)  \\
$54\,415\,174$		& -2087.549\,261	&	    -2087.711\,6(3)   &            -2087.709\,5(3)  \\
Extrap.             &    				&       -2087.734 		  &            -2087.738		\\
  \end{tabular}
\end{ruledtabular}
\end{table}
\end{squeezetable}

\begin{figure*}
	\begin{minipage}{0.35\linewidth}
		\includegraphics[width=\linewidth]{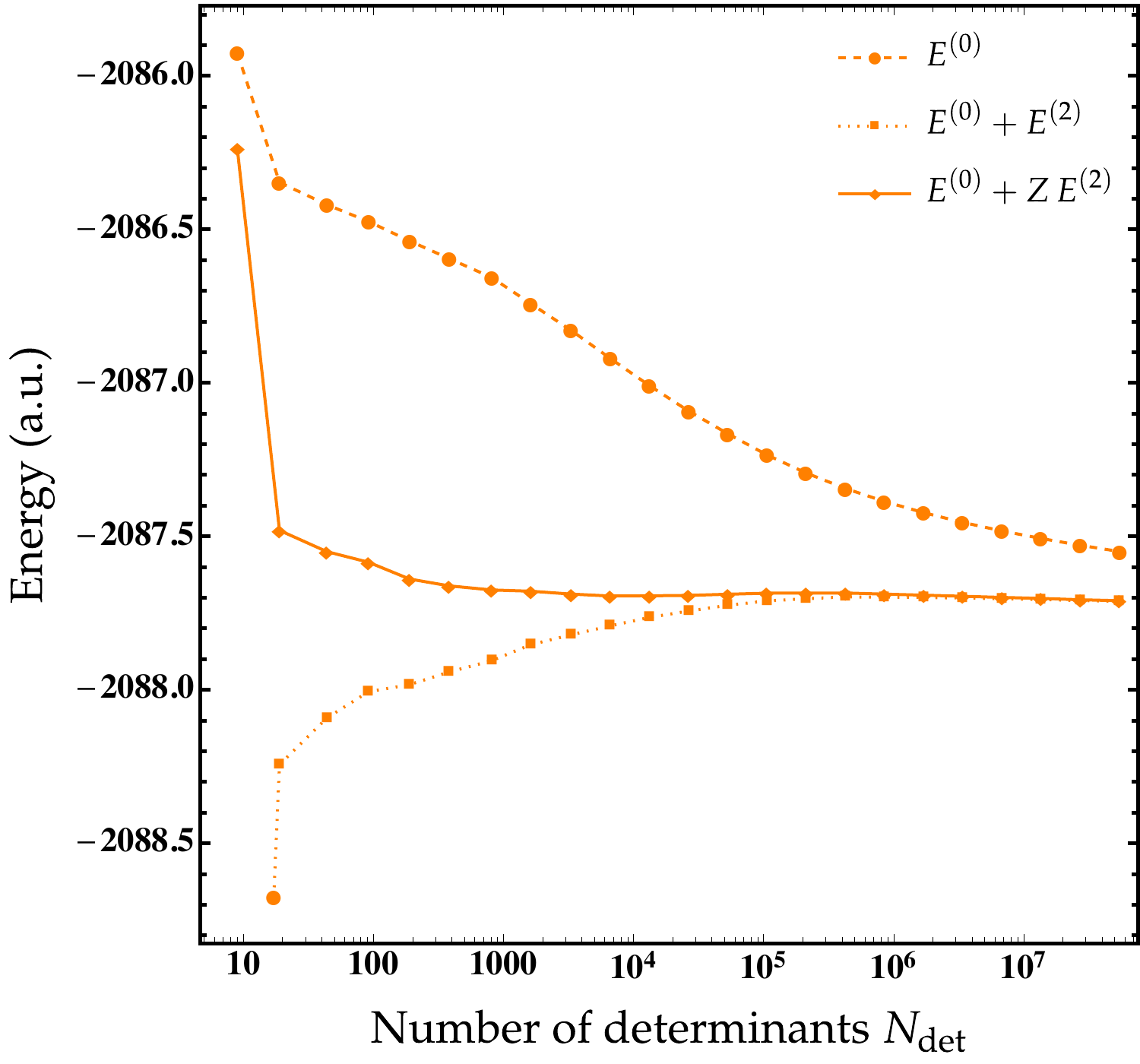}
		\includegraphics[width=\linewidth]{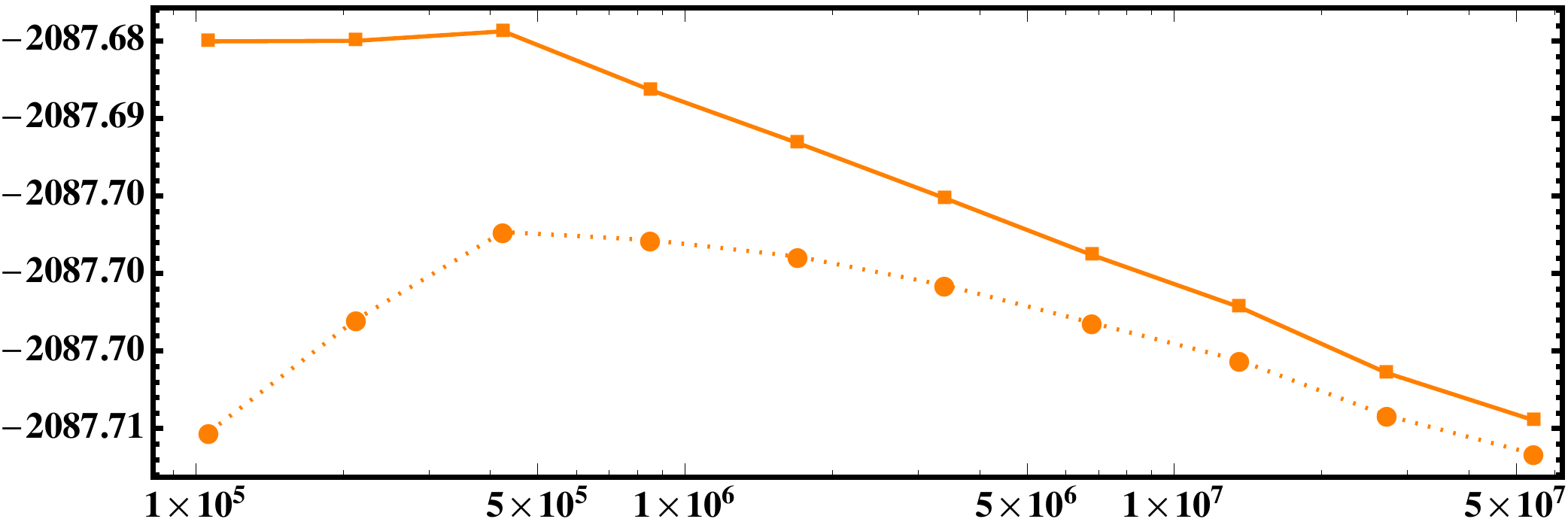}
	\end{minipage}
	\hspace{1cm}
	\begin{minipage}{0.45\linewidth}
		\includegraphics[width=\linewidth]{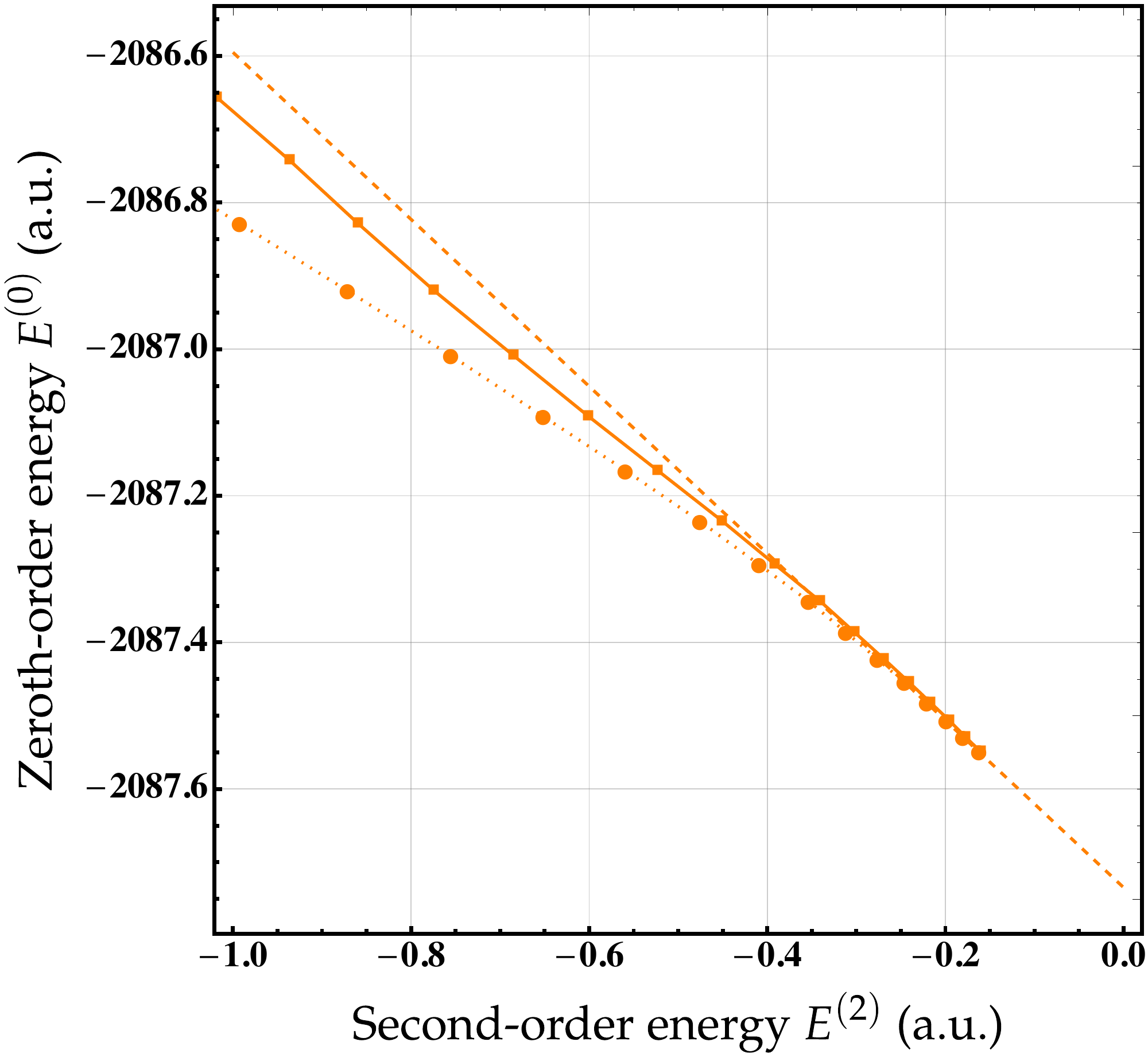}
	\end{minipage}
	\caption{Left: Energy convergence of the ground state of \ce{Cr2} with respect to the number of determinants $\Ndet$ in the reference space.
	The zeroth-order energy $\EO$ (dashed) , its second-order corrected value $\EO+\EPT$ (dotted) as well as its renormalized version $\EO+Z \EPT$ (solid) are represented.
	Right: Zeroth-order energy $\EO$ as a function of the second-order energy $\EPT$ (dotted lines) or its renormalization variant $Z\,\EPT$ (solid lines).
	A linear fit (dashed lines) of the last 2 points is also reported for comparison.
	See Table \ref{tab:Cr2} for raw data.}
	\label{fig:Cr2}
\end{figure*}

As a second test case for rPT2, we consider the widely-studied example of the chromium dimer (\ce{Cr2}) in its $^1 \Sigma_g^+$ ground state. \cite{Scuseria_1990,Roos_1995,Brynda_2009,Coe_2014,Purwanto_2015,Sokolov_2016,Sokolov_2017,Tsuchimochi_2017,LiManni_2013,Vancoillie_2016,Holmes_2016,Guo_2016,Sharma_2017,Garniron_2017b}
This system is notoriously challenging as it combines dynamic and static correlation effects hence requiring multi-configurational methods and large basis sets in order to have a balanced treatment of these two effects. 
Consequently, we compute its ground-state energy in the cc-pVQZ basis set with an internuclear distance $R_{\ce{Cr-Cr}} = 1.68~\AA$ close to its experimental equilibrium geometry.
Our full-valence calculation corresponds to an active space CAS(28,198) and the computational protocol is similar to the previous example.
The second-order corrected value $\EO+\EPT$ as well as its renormalized version $\EO+Z \EPT$ as a function of the number of determinants in the reference wave function are reported in Table \ref{tab:Cr2} and depicted in Fig.~\ref{fig:Cr2}.
Here also, we observe that rPT2 is clearly a superior extrapolation framework compared to the standard PT2 version as it yields a much straighter extrapolation curve, even in the case of a strongly correlated system such as \ce{Cr2}. 
The renormalization factor $Z$ [see Eq.~\eqref{eq:Z}] mitigates strongly the overestimation of the FCI energy for small wave functions by damping the second-order energy $\EPT$.
Linear extrapolations of the PT2 and rPT2 energies based on the two largest wave functions yields extrapolated FCI energies of -2087.734 and -2087.738, respectively (see also Table \ref{tab:Cr2}).
The difference between these two extrapolated FCI energies provides a qualitative idea of the extrapolation accuracy.

\subsection{Speedup}
\label{sec:perf}

In this Section, we discuss the parallel efficiency of the algorithms implemented in \QP. 
The system we chose for these numerical experiments is the benzene molecule \ce{C6H6} for which we have performed sCI calculations with the 6-31G* basis set.
The frozen-core approximation has been applied and the FCI space that we explore is a CAS(30,90).
The measurements were made on GENCI's Irene supercomputer. 
Each Irene's node is a dual-socket Intel(R) Xeon(R) Platinum 8168 CPU@2.70GHz with 192GiB of RAM, with a total of 48 physical CPU cores. 
Parallel speedup curves are made up to 12\,288~cores (i.e.~256 nodes) for i) a single iteration of the Davidson diagonalization, and ii) the hybrid semistochastic computation of $\EPT$ (which includes the CIPSI selection).
The speedup reference corresponds to the single node calculation (48 cores).

\begin{figure}
	\includegraphics[width=0.9\linewidth]{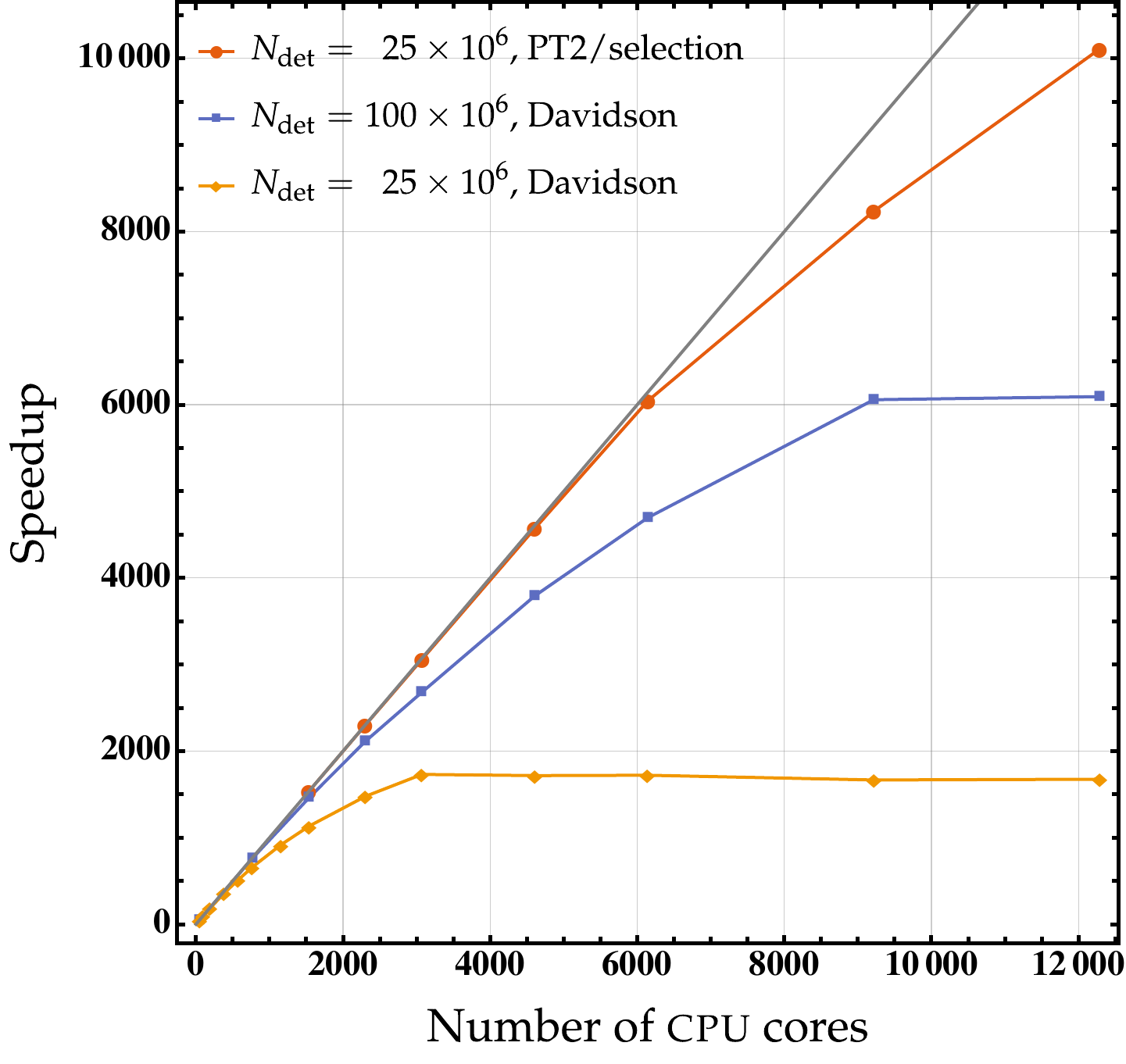}
	\caption{Speedup obtained for a single Davidson iteration (blue and yellow curves) and the combination of CIPSI selection and PT2 calculation (red curve) as a function of the number of CPU cores.
	For the Davidson diagonalization, two sizes of reference wave functions are reported ($\Ndet = 25 \times 10^6$ and $100 \times 10^6$), while for the PT2/selection calculation only results corresponding to the smallest wave function ($\Ndet = 25 \times 10^6$) are reported.
	See Table \ref{tab:speedup} for raw data.}
	\label{fig:speedup}
\end{figure}

First, we measure the time required to perform a single Davidson iteration as a function of the number of CPU cores for the two largest wave functions ($\Ndet = 25 \times 10^6$ and $100 \times 10^6$).
The timings are reported in Table \ref{tab:speedup} while the parallel speedup curve is represented in Fig.~\ref{fig:speedup}.
The parallel efficiency increases together with $\Ndet$, as shown in Fig.~\ref{fig:speedup}.
For the largest wave function, a parallel efficiency of 66\% is obtained on 192 nodes (i.e.~9216 cores).
We note that the speedup reaches a plateau at 3\,072 cores (64 nodes) for $\Ndet = 25 \times 10^6$.
For this wave function, there are 625 tasks computing each 40\,000 rows of $\mW$. 
When the number of nodes reaches 64, the number of tasks is too small for the load to be balanced between the nodes, and the computational time is limited by the time taken to compute the longest task. 
The same situation arises for $\Ndet = 100 \times 10^6$ with 9\,408 cores (192 nodes), with 2\,500 tasks to compute.

Second, we analyze the parallel efficiency of the calculation of $\EPT$ for the sCI wave function with $\Ndet = 25 \times 10^6$.
The stopping criterion during the calculation of $\EPT$ is given by a relative statistical error below $2 \times 10^{-3}$ of the current $\EPT$ value. 
The speedups are plotted in Fig.~\ref{fig:speedup} (see also Table \ref{tab:speedup}). 
For 192 nodes, one obtains a parallel efficiency of 89\%.
The present parallel efficiency is not as good as the one presented in the original paper.\cite{Garniron_2017}
The reason behind this is a faster computation of $\ePT{\alpha}$, which reduces the parallel efficiency by increasing the ratio communication/computation.

\begin{squeezetable}
\begin{table}
\caption{Wall-clock time (in seconds) to perform a single Davidson iteration and a second-order correction $\EPT$ calculation (which also includes the CIPSI selection) with an increasing number of 48-core compute nodes $\Nnodes$. 
The statistical error obtained on $\EPT$, defining the stopping criterion, is $0.17\times 10^{-3}$ a.u.
}
\label{tab:speedup}
\begin{ruledtabular}
\begin{tabular}{rddd}
\tabc{$\Nnodes$}	& \multicolumn{3}{c}{Wall-clock time (in seconds)} \\
		\cline{2-4}
		&	\tabc{Davidson for}		&	\tabc{Davidson for}		&	\tabc{PT2/selection}	\\
		&	\tabc{$\Ndet = 25 \times 10^6$}	&	\tabc{$\Ndet = 100 \times 10^6$}	&	\tabc{$\Ndet = 25\times 10^6$}			\\	
\hline
$1$		&	3\,340  	&	65\,915		&	406\,840	\\
$32$		&	142		&	2\,168		&	12\,711		\\
$48$		&	109		&	1\,497		&	8\,515		\\
$64$		&	93		&	1\,181		&	6\,421		\\
$96$		&	93		&	834		&	4\,323		\\
$128$		&	93		&	674		&	3\,287		\\
$192$		&	96		&	522		&	2\,435		\\
$256$		&	96		&	519		&	1\,996		\\
\end{tabular}
\end{ruledtabular}
\end{table}
\end{squeezetable}

\section{Developing in \QP}
\label{sec:dev}

\subsection{The \QP philosophy}
\label{sec:philo}
\QP is a standalone easy-to-use library for developers.
The main goals of \QP are to
i) facilitate the development of new quantum chemistry methods,
ii) minimize the dependency on external {programs/libraries}, and
iii) encourage the collaborative and educative work through human readable programs.
Therefore, from the developer point of view, \QP can be seen as a standalone library containing all important quantities needed to perform quantum chemistry calculations, both involving wave function theory, through the determinant driven algorithms, and DFT methods, thanks to the presence of a quadrature grid for numerical integrations and basic functionals. 
These appealing features are made more concrete thanks to the organization of \QP in terms of core modules and plugins (see Sec.~\ref{sec:plugins}) together with its programming language (see Sec.~\ref{sec:irpf90}), which naturally creates a very modular environment for the programmer. 

Although \QP is able to perform all the required steps from the calculation of the one- and two-electron integrals to the computation of the sCI energy, interfacing \QP, at any stage, with other programs is relatively simple.
For example, canonical or CASSCF molecular orbitals can be imported from GAMESS, \cite{gamess} while atomic and/or molecular integrals can be read from text files like \textsc{fcidump}.
Thanks to this flexibility, some of us are currently developing plugins for performing sCI calculations for periodic systems.

\subsection{The IRPF90 code generator}
\label{sec:irpf90}

It is not a secret that large scientific codes written in Fortran (or in similar languages) are difficult to maintain.
The program's complexity originates from the inter-dependencies between the various entities of the code.
As the variables are more and more coupled, the programs become more and more difficult to maintain and to debug.
To keep a program under control, the programmer has to be aware of all the consequences of any source code modification within all possible execution paths.
When the code is large and written by multiple developers, this becomes almost impossible.
However, a computer can easily handle such a complexity by taking care of all the dependencies between the variables, in a way similar to how GNU Make handles the dependencies between source files.

IRPF90 is a Fortran code generator. \cite{Scemama_2009} 
Schematically, the programmer only writes computation kernels, and IRPF90 generates the glue code linking all these kernels together to produce the expected result, handling all relationships between variables.
To illustrate in a few words how IRPF90 works, let us consider the simple example which consists of calculating the total energy of a molecular system as the sum of the nuclear repulsion and the electronic energy $E_\text{tot} = E_\text{nuc} + E_\text{ele}$.
The electronic energy is the sum of the kinetic and potential energies, i.e., $E_\text{ele} = E_\text{kin} + E_\text{pot}$.

\begin{figure}
	\includegraphics{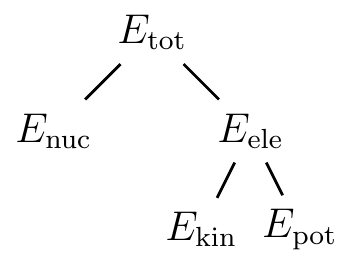}
	\caption{Production tree of the energy computed by IRPF90.}
	\label{fig:irp_tree}
\end{figure}

The \emph{production tree} associated with the computation of the total energy is shown in Fig.~\ref{fig:irp_tree}.
Within the IRPF90 framework, the programmer writes a \emph{provider} for each \emph{entity}, i.e., a node of the production tree.
The provider is a subroutine whose only goal is to compute the value associated with the entity, assuming the values of the entities on which it depends are computed and valid.
Hence, when an entity is used somewhere in the program (in a subroutine, a function or a provider), a call to its provider is inserted in the code before it is used such that the corresponding value is guaranteed to be valid.

\QP is a library of providers designed to make the development of new wave function theory and DFT methods simple.
Only a few programs using these providers are part of the core modules of \QP, such as the sCI module using the CIPSI algorithm or the module containing the semi-stochastic implementation of the second-order perturbative correction.
The main goal of \QP is to be used as a library of providers, and programmers are encouraged to develop their own modules using \QP.

\subsection{The plugin system}
\label{sec:plugins}

External programmers should not add their contributions by modifying directly \QP's core, but by creating their own modules in independent repositories hosted and distributed by themselves.
This model gives more freedom to the developers to distribute modules as we do not enforce them to follow any rule.
The developers are entirely responsible for their own plugins.
This model has the advantage to redirect immediately the users to the right developer for questions, installation problems, bug reports, etc.

\QP integrates commands to download external repositories and integrate all the plugins of these repositories into the current installation of \QP.
External plugins appear exactly as if they were part of \QP, and if a plugin is useful for many users, it can be easily integrated in \QP's core after all the coding and documentation standards are respected.

Multiple external plugins were developed by the authors.
For instance, one can find a multi-reference coupled cluster program,\cite{Giner_2016, Garniron_2017} interfaces with the quantum Monte Carlo programs QMC=Chem,\cite{Scemama_2013b} QMCPack\cite{Kim_2018} and CHAMP,\cite{champ} an implementation of the shifted-Bk method,\cite{Garniron_2018} a program combining CIPSI with RSDFT,\cite{Giner_2018} a four-component relativistic RSDFT code, \cite{Paquier_2018} and many others.

In particular, \QP also contains the basic tools to use and develop range-separated density-functional theory (RSDFT, see, e.g., Refs.~\onlinecite{Sav-INC-96,TouColSav-PRA-04}) which allows to perform multi-configurational density-functional theory (DFT) calculations within a rigorous mathematical framework. 
In the core modules of \QP, single-determinant approximations of RSDFT are available, which fall into the so-called range-separated hybrid \cite{janosh_rsh_05,AngGerSavTou-PRA-05} (RSH) approximation. These approaches correct for the wrong long-range behavior of the usual hybrid approximations thanks to the inclusion of the long-range part of the HF exchange.  
\QP contains all necessary integrals to perform RSDFT calculations, including the long-range interaction integrals and Hartree-exchange-correlation energies and potentials derived from the short-range version of the local-density approximation (LDA) \cite{PazMorGorBac-PRB-06} and a short-range generalized-gradient approximation (GGA) based on the Perdew-Burke-Ernzerhof (PBE) functional. \cite{GolWerStoLeiGorSav-CP-06}
All numerical integrals are performed using the standard Becke quadrature grid \cite{Becke_1988} associated with the improved radial grids of Mura et \emph{al.}\cite{Knowles_1996}
With these tools, more evolved schemes based on RSDFT have been developed, such as an energy correlation functional with multideterminantal reference depending on the on-top pair density \cite{FerGinTou-JCP-18} or a basis set correction. \cite{Giner_2018}
The corresponding source code can be found as external plugins (see, for example, \url{https://gitlab.com/eginer/qp_plugins_eginer}).

\section{Conclusion}
\label{sec:ccl}
Significant improvements were brought to the second version of \QP. 
Some were single-core optimizations, and others focused on the algorithm adaptation to large-scale parallelism (load balancing in particular). 
Currently, the code has a parallel efficiency that enables routinely to realize runs on roughly 2\,000 CPU cores, with tens of millions of determinants in the reference space. 
Moreover, we have been able to push up to 12\,288 cores (256 nodes) on GENCI's supercomputer Irene. 
Such a gain in efficiency has and will lead to many more challenging chemical applications. \cite{Caffarel_2014, Giner_2013, Giner_2015, Caffarel_2016a, Caffarel_2016b, Loos_2018b, Loos_2019, Scemama_2018b, Dash_2018, Flores_2018}

The Davidson diagonalization, which is at the center of sCI and FCI methods, suffers from the impossibility to fully store the Hamiltonian in the memory of a single node. The solution we adopted was to resort to \emph{direct methods}, i.e., recomputing \emph{on the fly} the matrix elements at each iteration. 
While an extremely fast method was already available to detect zero matrix elements,\cite{Scemama_2013a} the former implementation still had to search over the $\order*{\Ndet^2}$ matrix elements for interacting determinant pairs. 
Now, determinants are split in disjoint sets entirely disconnected from each other. 
Thus, only a small fraction of the matrix elements need to be explored, and an algorithm with $\order*{\Ndet^{3/2}}$ scaling was proposed. 
While the parallelization of this method was somewhat challenging due to the extremely unbalanced nature of the elementary tasks, a distributed implementation was realized with satisfying parallel speedups (typically 35 for 50 nodes) with respect to the 48-core single-node reference.

Significant improvements were also realized in the computation of the second-order perturbative correction, $\EPT$. 
A natural idea was to take into account the tremendous number of tiny contributions via a stochastic Monte Carlo approach.
$\EPT$ being itself an approximate quantity used for estimating the FCI energy, its exact value is indeed not required, as long as the value is unbiased and the statistical error is kept under control.
Our scheme allows to compute $\EPT$ with a small error bar for a few percent of the cost of the fully deterministic computation.  

Similarly, the CIPSI selection is now performed stochastically alongside the PT2 calculation.
Therefore, the selection part of the new stochastic CIPSI selection is virtually free as long as one is interested in the second-order perturbative correction.

Finally, efforts have been made to make this software as developer friendly as possible thanks to a very modular architecture that allows any developer to create his/her own module and to directly benefit from all pre-existing work. 

\section*{License}
\label{sec:license}
\QP is licensed under GNU Affero General Public License (AGPLv3).

\begin{acknowledgements}
The authors would like to thank the \emph{Centre National de la Recherche Scientifique} (CNRS) for funding and Cyrus Umrigar for carefully reading the manuscript.
Funding from \emph{Projet International de Coop\'eration Scientifique} (PICS08310) is also acknowledged.
This work was performed using HPC resources from CALMIP (Toulouse) under allocation 2019-18005 and from GENCI-TGCC (Grant 2018-A0040801738).
A.B.~was supported by the U.S.~Department of Energy, Office of Science, Basic Energy Sciences, Materials Sciences and Engineering Division, as part of the Computational Materials Sciences Program and Center for Predictive Simulation of Functional Materials.
K.G.~acknowledges support from grant number CHE1762337 from the U.S.~National Science Foundation.
\end{acknowledgements}

\appendix

\section{Implementation details}
\label{sec:imple_details}

\subsection{Efficiency of integral storage}
\label{sec:imple_details_eri}

The efficiency of the storage as a hash table was measured on a dual socket Intel Xeon E5-2680 v2@2.80GHz processor, taking the water molecule with the cc-pVQZ basis set (115 MOs). 
The time to access all the integrals was measured by looping over the entire set of ERIs using different loop orderings. 
The results are given in Table~\ref{tab:integrals}, the reference being the storage as a plain four-dimensional array.

\begin{table}
\caption{Time to access integrals (in nanoseconds/integral) with different access patterns. 
The time to generate random numbers (measured as 67~ns/integral) was not counted in the random access results.}
\label{tab:integrals}
  \begin{ruledtabular}
  \begin{tabular}{ldd}
   Access  & \tabc{Array}  & \tabc{Hash table}  \\ 
\hline
 $i,j,k,l$ &  9.72  &  125.79     \\ 
 $i,j,l,k$ &  9.72  &  120.64     \\ 
 $i,k,j,l$ & 10.29  &  144.65     \\ 
 $l,k,j,i$ & 88.62  &  125.79     \\ 
 $l,k,i,j$ & 88.62  &  120.64     \\ 
   Random  & 170.00 &  370.00     \\ 
  \end{tabular}
  \end{ruledtabular}
\end{table}

In the array storage, the value of 170~ns/integral in the random access case is typical of the latency to fetch a value in the RAM modules, telling that the requested integral is almost never present in any level of cache.
When the data is accessed with a stride of one ($i,j,l,k$ storage), the hierarchical architecture of the cache levels accelerates the access by a factor of 18, down to 9.71~ns/integral, corresponding mostly to the overhead of the function call, the retrieval of the data being negligible.

With the hash table, the random access is only 2.18 times slower than the random access in the array. 
Indeed, two random accesses are required: one for the first element of the key bucket to do the search, and one for the value of the integral. 
The remaining time corresponds to the binary search.
The results show that data locality is exploited: when the access is done with a regular access pattern, the data is fetched roughly 3 times faster than using a random access, giving a latency below the latency of a random access in the array.

A CIPSI calculation was run once with the array storage, and once with the hash table storage. 
With the hash storage, the total wall clock time was increased only by a factor of two.
To accelerate the access to the most frequently used integrals and reduce this overhead, we have implemented a software cache. 
All the integrals involving the 128 MOs closest to the Fermi level are copied in a dense array of 128$^4$ elements (2~GiB), and benefit from the fastest possible access.

\subsection{Internal representation of determinants}
\label{sec:imple_details_det}

Determinants can be conveniently written as a string of creation operators applied to the vacuum state $\vac$, e.g., $\ac{i} \ac{j} \ac{k} \vac = \kI$.
Because of the fermionic nature of electrons, a permutation of two contiguous creation operators results in a sign change $\ac{j} \ac{i} = -\ac{i} \ac{j}$, which makes their ordering relevant, e.g., $\ac{j} \ac{i} \ac{k} \vac =  -\kI$.
A determinant can be broken down into two pieces of information:
i) a set of creation operators corresponding to the set of occupied spinorbitals in the determinant, and ii) an ordering of the creation operators responsible for the sign of the determinant, known as \emph{phase factor}.
Once an ordering operator $\ordering$ is chosen and applied to all determinants, the phase factor may simply be included in the CI coefficient.

The determinants are built using the following order: i) spin-up ($\uparrow$) spinorbitals are placed before spin-down ($\downarrow$) spinorbitals, as in the Waller-Hartree double determinant representation\cite{Pauncz_1989} $\ordering \kI = \hI \vac = \hI_\uparrow \hI_\downarrow \vac$, and ii) within each operator $\hat{I}_\uparrow$ and $\hat{I}_\downarrow$, the creation operators are sorted by increasing indices.
For instance, let us consider the determinant $\kJ = \ac{j} \ac{k} \ac{\bar i} \ac{i} \vac$ built from the set of spinorbitals $\{i_{\uparrow},j_{\uparrow},k_{\uparrow},i_{\downarrow} \}$ with $i<j<k$.
If we happen to encounter such a determinant, our choice of representation imposes to consider its re-ordered expression $\ordering \kJ = - \ac{i} \ac{j} \ac{k} \ac{\bar i} \vac = -\kJ$, and the phase factor must be handled.

The indices of the creation operators (or equivalently the spinorbital occupations), are stored using the so-called \emph{bitstring} encoding. 
A bitstring is an array of bits; typically, the 64-bit binary representation of an integer is a bitstring of size 64.
Quite simply, the idea is to map each spinorbital to a single bit with value set to its occupation number. 
In other words, 0 and 1 are associated with the \emph{unoccupied} and \emph{occupied} states, respectively.

For simplicity and performance considerations, the occupations of the spin-up and spin-down spinorbitals are stored in different bitstrings, rather than interleaved or otherwise merged in the same one. 
This allows to straightforwardly map orbital index $p$ to bit index $p-1$ (orbitals are usually indexed from 1, while bits are indexed from 0).
This makes the representation of a determinant a tuple of two bitstrings, associated with respectively spin-up and spin-down orbitals. 
A similar parity representation of the fermionic operators is commonly used in quantum computing. \cite{Seeley_2012}

The storage required for a single determinant is, in principle, one bit per spinorbital, or $2 \times \Norb$ bits. However, because CPUs are designed to handle efficiently 64-bit integers, each spin part is stored as an array of 64-bit integers, the unused space being padded with zeros. 
The actual storage needed for a determinant is $2 \times 64 \times \Nint$ bits, where $\Nint = \left \lfloor (\Norb-1)/64 \right \rfloor + 1$ is the number of 64-bit integers needed to store one spin part.

Taking advantage of low-level hardware instructions, \cite{Scemama_2013a} we are able, given two arbitrary determinants $\kI$ and $\kJ$, to find with a minimal cost the excitation operator $\hat{T}$ such that $\kJ = \hat{T} \kI$. This is a necessary step to obtain the $(i,j,k,l)$ indices of the two-electron integral(s) involved in the Hamiltonian matrix element between $\kI$ and $\kJ$.
Then, fetching the values of the integrals can be done quickly using the hash table presented in Sec.~\ref{sec:detdrive}.

Because the data structure used to store determinants implies an ordering of the MOs, we also need to compute a phase factor.
Here, we propose an algorithm to perform efficiently the computation of the phase factor. 
For a determinant $\kI$ that is going to be used repeatedly for phase calculations, we introduce a \emph{phase mask} represented as a bitstring:
\begin{equation}
	P_I[i] = 1 \wedge \sum_{k=0}^i I[k],
\end{equation}
where $\wedge$ denotes the \emph{and} bitwise operation, and $I[k]$ is the $k$th bit of bitstring $I$, corresponding to the $(k+1)$th spinorbital of determinant $\kI$ (remember that the orbital indices start at 1 and the bit indices start at 0).
In other words, the $i$th bit of the phase mask is set to 1 if the number of electrons occupying the $i+1$ lowest spinorbitals is odd, and 0 otherwise.
When an electron of determinant $\kI$ is excited from orbital $h$ to $p$, the associated phase factor is 
\begin{equation}
\begin{cases}
	+(-1)^{P_I[h-1] \oplus P_I[p-1]}, & \text{if } p>h, \\
	-(-1)^{P_I[h-1] \oplus P_I[p-1]}, & \text{if } h>p,
\end{cases}
\end{equation}
where $\oplus$ denotes the exclusive or (\emph{xor}) operation. 
So if the phase mask is available, the computation of the phase factor only takes a few CPU cycles.
Another important aspect is to create efficiently the phase masks. We propose Algorithm~\ref{alg:PHASEMASK}, which computes it in a logarithmic time for groups of 64 MOs, taking advantage of the associativity of the exclusive \textit{or} operator. 
Indeed, the ``for'' loop executes 6 cycles to update the mask for $2^6=64$ MOs.

\begin{algorithm}
\caption{Function that returns a phase mask as a bitstring.}
  \label{alg:PHASEMASK}
  \SetKwFunction{FMain}{PhasemaskOfDet}
  \SetKwProg{Fn}{Function}{:}{}

  \Fn(){\FMain{$\bitI$}}{
    \KwData{ $\bitI$ : 64-bit string representation of $\ket{I}$}
    \KwResult{ $\bitP$ : phase mask associated with $\ket{I}$, as a 64-bit string.}

    \For{$\sigma \in \{\uparrow, \downarrow\}$}{
    $r \gets 0$ \;
    \For{$i \gets 0, \Nint-1$}{
        $\bitPsigma[i] \gets \ieor{\bitIsigma[i]}{\shiftl{\bitIsigma[i]}{1}}$ \;
        \For{$d \gets 0, 5$}{
                $\bitPsigma[i] \gets \ieor{\bitPsigma[i]}{\shiftl{P_\sigma[i]}{\shiftl{1}{d}}}$ \;
        }
        $\bitPsigma[i] \gets \ieor{\bitPsigma[i]}{r}$ \;
        \If{$\iand{\qty(\popcnt{\bitIsigma[i]}}{1}) = 1$}{
                $r \gets \neg{r}$ \;
        }
      }
    }
    \KwRet{$P$} \;
  } 
\begin{tabular}{ll}
\hline
$\popcnt{I}$    &: number of bits set to 1 in $I$ (\emph{popcnt}), \\
$\wedge$        &: bitwise \emph{and}, \\
$\oplus$        &: bitwise \emph{xor}, \\
$\shiftl{I}{k}$ &: shift $I$ by $k$ bits to the left, \\
$\neg$          &: bitwise negation. 
\end{tabular}
\end{algorithm}

\subsection{Davidson diagonalization}
\label{sec:imple_details_dav}

Within \QP, the Davidson diagonalization algorithm is implemented in its multi-state version. 
Algorithmically, the expensive part of the Davidson diagonalization is the computation of the matrix product $\mH\,\mU$.
As mentioned above (see Sec.~\ref{sec:meth}), two determinants $\kI$ and $\kJ$ are connected via $\mH$ (i.e.~$\mel*{I}{\hH}{J} \neq 0$) only if they differ by no more than two spinorbitals. 
Therefore, the number of non-zero elements per row in $\mH$ is equal to the number of single and double excitation operators, namely $\order*{\Nalpha^2 (\Norb - \Nalpha)^2}$. 
As $\mH$ is symmetric, the number of non-zero elements per column is identical. 
This makes $\mH$ very sparse. 
However, for large basis sets, the whole matrix may still not fit in a single node memory, as the number of non-zero entries to be stored is of the order of $\Ndet \Nalpha^2 (\Norb - \Nalpha)^2$.  
One possibility would be to distribute the storage of $\mH$ among multiple compute nodes, and use a distributed library such as PBLAS\cite{pblas} to perform the matrix-vector operations. 
Another approach is to use a so-called \emph{direct} algorithm, where the matrix elements are computed \emph{on the fly}, and this is the approach we have chosen in \QP.
This effectively means iterating over all pairs of determinants $\kI$ and $\kJ$, checking whether $\kI$ and $\kJ$ are connected by $\mH$ and if they are, accessing the corresponding integral(s) and computing the phase factor. 
Even though it is possible to compute the excitation degree between two determinants very efficiently,\cite{Scemama_2013a} the number of such computations scales as $\Ndet^2$, which becomes rapidly prohibitively high.
To get an efficient determinant-driven implementation it is mandatory to filter out all pairs of determinants that are not connected by $\mH$, and iterate only over connected pairs. 
To reach this goal, we have implemented an algorithm similar to the \emph{Direct Selected Configuration Interaction Using Strings} (DISCIUS) algorithm.\cite{Povill_1995}

The determinants of the internal space are re-ordered in linear time as explained in Ref.~\onlinecite{Scemama_2016}, such that the wave function can be expressed as
\begin{equation}
\label{eq:sce2016}
	\ket*{\PsiO}  = \sum_{I}^{\Nalphadet} \sum_{J}^{\Nbetadet} C_{IJ} \ket{I_\uparrow J_\downarrow},
\end{equation}
where we take advantage of the Waller-Hartree double determinant representation. \cite{Pauncz_1989} 

Moving along a row or a column of $\mC$ keeps the spin-up or spin-down determinants fixed, respectively. 
For a given determinant, finding the entire list of same-spin single and double excitations can be performed in $\order*{\Nalphadet} = \order*{\Nbetadet} =\order*{\sqrt{\Ndet}}$, while finding the opposite-spin double excitations is done via a two-step procedure.
First, we look for all the spin-up single excitations. 
Then, starting from this list of spin-up single excitations, we search for the spin-down single excitation such that the resulting opposite-spin doubly-excited determinant belongs to $\PsiO$.
Hence, the formal scaling is reduced to $\order*{\Ndet^{3/2}}$.
It could be further reduced to $\order*{\Ndet}$ at the cost of storing the list of all singly- and doubly-excited determinants for each spin-up and spin-down determinant, but we preferred not to follow this path in order to reduce the memory footprint as much as possible.

\subsection{CIPSI selection and PT2 energy}
\label{sec:imple_details_sel}

There are multiple ways to compute the $\ePT{\alpha}$'s.
One way is to loop over pairs of internal determinants $\kI$ and $\kJ$, generate the list of external determinants $\qty{\kalpha}$ connecting $\kI$ and $\kJ$ and increment the corresponding values $\ePT{\alpha}$ stored in a hash table.
Using a hash table to store in memory a list of $\kalpha$'s without duplicates and their contributions $\ePT{\alpha}$ is obviously not a reasonable choice since the total number of $\kalpha$'s scales as $\order*{\Ndet \Nalpha^2 \qty(\Norb - \Nalpha)^2}$.
To keep the memory growth in check, we must design a function that can build a stream of unique external determinants, compute their contribution $\ePT{\alpha}$ and retain in memory only the few most significant pairs $(\kalpha,\ePT{\alpha})$.

In {\QP}, we build the stream of unique external determinants as follows.
We loop over the list of internal determinants (the \emph{generators}) sorted by decreasing $c_I^2$.
For each generator $\kI$, we generate all the singly- and doubly-excited determinants $\qty{\kalpha}$, removing from this set the internal determinants and the determinants connected to any other generator $\kJ$ such that $J < I$.
This guarantees that the $\kalpha$'s are considered only once, without any additional memory requirement.

For each generator $\kI$, before generating its set of $\kalpha$'s, we pre-compute the diagonal of the Fock matrix associated with $\kI$.
This enables to compute the diagonal elements $\Hij{\alpha}{\alpha}$ involved in Eq.~\eqref{eq:e2} for a few flops.\cite{Cimiraglia_1996}
The computation of $\Hij{\PsiO}{\alpha} = \sum_J c_J \Hij{J}{\alpha}$ is more challenging than the diagonal term since, at first sight, it appears to involve the $\Ndet$ internal determinants. 
Fortunately, most of the terms amongst this sum vanish due to Slater-Condon's rules. 
Indeed, we know that the terms where $\kJ$ is more than doubly excited with respect to $\kalpha$ vanish, and these correspond to the determinants $\kJ$ which are more than quadruply excited with respect to $\kI$. \cite{Cimiraglia_1996}
To compute efficiently $\Hij{\PsiO}{\alpha}$, for each generator $\kI$, we create a filtered wave function $\ket*{\PsiO_I}$ 
by projecting $\ket*{\PsiO}$ on a subset $\mathcal{J}_I$ of internal determinants $\{\kJ\}$ where $\Hij{J}{\alpha}$ is possibly non-zero. 
This yields $\Hij{\PsiO}{\alpha} = \Hij{\PsiO_I}{\alpha}$, where $\PsiO_I$ is a much smaller determinant expansion than $\PsiO$.
In addition, as we have defined the $\kalpha$'s in such a way that they do not interact with $\kJ$ when $J < I$, all these $\kJ$'s can also be excluded from $\mathcal{J}_I$.
This pruning process yielding to $\ket*{\PsiO_I}$ will be referred to as the \emph{coarse-grained} filtering.
A \emph{fine-grained} filtering of $\ket*{\PsiO_I}$ is performed in a second stage to reduce even more the number of determinants, as we shall explain later.

To make the coarse-grained filtering efficient, we first filter out the determinants that are more than quadruply excited in the spin-up and spin-down sectors separately.
Using the representation shown in Eq.~\eqref{eq:sce2016}, this filtering does not need to run through all the internal determinants and scales as $\order*{\Ndet^{\uparrow}} = \order*{\sqrt{\Ndet}}$.
It is important to notice that, at this stage, the size of $\mathcal{J}_I$ is bounded by the number of possible quadruple excitations in both spin sectors, and does not scale any more as $\order*{\Ndet}$.
Next, we remove the determinants that are i) quadruply excited in one spin sector and excited in the other spin sector, ii) triply excited in one spin sector and more than singly excited in the other spin sector, and iii) all the determinants that are doubly excited in one spin sector and more than doubly excited in the other spin sector.

The external determinant contributions are computed in batches.
A batch $I_{pq}$ is defined by a doubly-ionized generator $\ket{I_{pq}} = a_p a_q \kI$.
When a batch is created, the fine-grained filtering step is applied to $\mathcal{J}_I$ to produce $\mathcal{J}_{I_{pq}}$ and  $\PsiO_{I_{pq}}$, such that $\Hij{\PsiO_{I_{pq}}}{\alpha} = \Hij{\PsiO_I}{\alpha}$.

Each external determinant produced in the batch $I_{pq}$ is characterized by two indices $r$ and $s$ with $\ordering a^\dagger_r a^\dagger_s a_p a_q  \ket*{I} = \kIpqrs$. The contribution associated with each determinant of a given batch will be computed incrementally in a two-dimensional array $A(r,s)$ as follows.
A first loop is performed over all the determinants $\kJ$ belonging to the filtered internal space $\mathcal{J}_{I_{pq}}$.
Comparing $\kJ$ to $\ket{I_{pq}}$ allows to quickly identify if $\kJ$ will be present in the list of external determinants, and consequently tag the corresponding cell $A(r,s)$ as \emph{banned}.
Banned cells will not be considered for the computation of $\ePT{\alpha}$ nor the determinant selection, as they correspond to determinants already belonging to the internal space.
A second loop over all the $\kJ \in \mathcal{J}_{I_{pq}}$ is then performed. 
During this loop, all the $(r,s)$ pairs where $\kIpqrs$ is connected to $\kJ$ are generated, and the corresponding cells $A(r,s)$ are incremented with $c_J \Hij{J}{\Ipqrs}$.
After this second loop, $A(r,s) = \Hij{\Psi}{\Ipqrs}$ and all the contributions $\ePT{\alpha}$ of the batch can be obtained using $A(r,s)$.
The running value of $\EPT$ is then incremented, and the $\Ndet$ most significant determinants are kept in an array sorted by decreasing $\abs*{\ePT{\alpha}}$.

\begin{figure}
	\includegraphics[width=\linewidth]{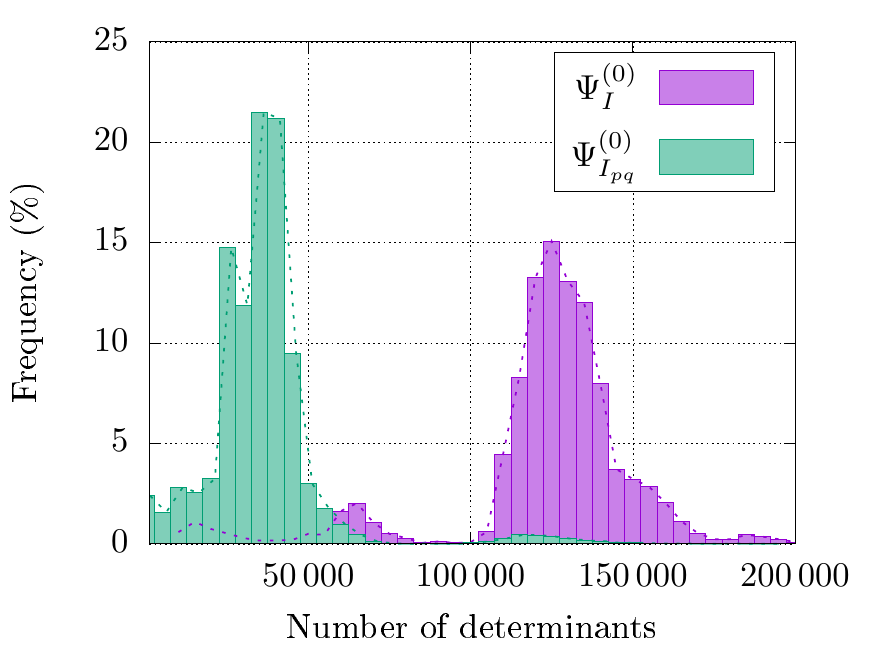}
	\caption{Histograms representing the number of determinants remaining after the coarse-grained (purple) and fine-grained (green) filtering processes applied to the ground state of the CN3 molecule with $\Ndet = 935\,522$.}
	\label{fig:filtering}
\end{figure}

Figure~\ref{fig:filtering} shows the number of determinants retained in $\PsiO_I$ or $\PsiO_{I_{pq}}$ after filtering out disconnected determinants of the ground state of the CN3 molecule with 935\,522 determinants (see Sec.~\ref{sec:imple_details}).
This example shows that, starting from $\PsiO$, the coarse-grained process which consists of removing the determinants more than quadruply excited with respect to the generator $\kI$ produces wave functions $\PsiO_I$ with a typical size of 120\,000 determinants, a reduction by a factor 8.
Then, starting from $\PsiO_I$, the fine-grained filtering, specific to the batch generating $\PsiO_{I_{pq}}$, reduces even more the number of determinants (by a factor 3), down to a typical size of 40\,000 determinants, which represents only $4\%$ of the total wave function $\PsiO$.

\bibliography{QP2,QP2-control}

\end{document}